\UseRawInputEncoding
\documentclass[11pt]{article}
\usepackage[table]{xcolor}
\usepackage{float}
\usepackage{pifont}
\usepackage{rotating}
\usepackage[normalem]{ulem}
\usepackage[numbers]{natbib}
\usepackage[table]{xcolor}
\usepackage{subfigure}
\usepackage{tikz}
\usepackage{amsmath,amssymb,amsmath,amsthm,epsfig,graphicx,natbib}
\usepackage{color}
\usepackage{appendix}
\usepackage{enumitem}
\usepackage{authblk}
\usepackage{lscape}
\usepackage{bm}
\usepackage{lineno,hyperref}
\usepackage{algorithm}
\usepackage{algorithmicx}
\usepackage{algpseudocode}
\usepackage{mwe}
\usepackage{colortbl}
\usepackage{multirow}
\usepackage{appendix}
\usepackage{soul}
\usepackage{caption}
\usepackage{stackrel}
\usepackage{enumerate}
\usepackage{url} 
\usepackage{amsfonts}

\usepackage{hyperref}
\hypersetup{%
  colorlinks=true,
  linkcolor=blue,
  citecolor=blue,
}

\newtheorem{prop}{Proposition}

\newtheorem{lem}{Lemma}
\newtheorem{thm}{Theorem}
\newtheorem{ass}{Assumption}
\newtheorem{axi}{Axiom}
\newtheorem{ex}{Example}
\newtheorem{counterex}{Counterexample}
\newtheorem{rem}{Remark}
\newtheorem*{ack}{Acknowledgement}
\newenvironment{proposition}{\begin{prop}\hspace{-2mm}}{\end{prop}}

\newcommand{\Keywords}[1]{\par\noindent
{\small{\bf Keywords\/}: #1}}

\begin{document}


\title{A cost-sensitive constrained Lasso}

\author[1,2]{Rafael Blanquero}
\author[1,2]{Emilio Carrizosa}
\author[1,3]{Pepa Ram\'irez-Cobo}
\author[1,2]{M. Remedios Sillero-Denamiel}

\affil[1]{Instituto de Matem\'aticas de la Universidad de Sevilla (IMUS), Seville, Spain}
\affil[2]{ Departamento de Estad\'istica e Investigaci\'on Operativa, Universidad de Sevilla.}
\affil[3]{Departamento de Estad\'istica e Investigaci\'on Operativa, Universidad de C\'adiz.}

\date{}

\maketitle

\begin{abstract}
\noindent
The Lasso has become a benchmark data analysis procedure, and numerous variants have been proposed in the literature. Although the Lasso formulations are stated so that overall prediction error is optimized, no full control over the accuracy prediction on certain individuals of interest is allowed. 

In this work we propose a novel version of the Lasso in which quadratic performance constraints are added to Lasso-based objective functions, in such a way that threshold values are set to bound the prediction errors in the different groups of interest (not necessarily disjoint). As a result, a constrained sparse regression model is defined by a nonlinear optimization problem. This cost-sensitive constrained Lasso has a direct application in heterogeneous samples where data are collected from distinct sources, as it is standard in many biomedical contexts. Both theoretical properties and empirical studies concerning the new method are explored in this paper. In addition, two illustrations of the method on biomedical and sociological contexts are considered.
\vspace*{0.1cm}

\noindent
\end{abstract}

\section{Introduction}\label{introduction}

Let $(Y,X)$ be a random vector, where $X = (X_1,\ldots,X_p)$ is a vector of $p$ predictors and $Y$ identifies the response variable. Given the observed response vector $\mathbf{y}= (y_1, \ldots, y_n)'$, $n>p$, and the related observed predictors, $\mathbf{x}_j=(x_{1j},\ldots,x_{nj})'$, $j=1,\ldots,p$, the linear regression model predicts $\mathbf{y}$ by

\begin{equation*}
\mathbf{\hat{y}}= \hat{\beta}_0 + \hat{\beta}_1 \mathbf{x}_1 + \ldots + \hat{\beta}_p \mathbf{x}_p.
\end{equation*}

Consider the well-known {\tt prostate} database \cite{prostate}, which consists of the measurements of $p=8$ predictors and one response variable (clinical measures) on $n=97$ men who were about to receive a radical prostatectomy. Further, assume that the dataset is divided into two groups: \emph{Group 1}, corresponding to \emph{young} individuals (aged less than 65) and \emph{Group 2}, related to the population older than 65. If the goal is to minimize the overall mean squared error (MSE), the parameter vector $\boldsymbol{\beta}=(\beta_0, \beta_1, \ldots, \beta_p)$ can be estimated by a fitting procedure as ordinary least squares (OLS), yielding $\boldsymbol{\hat{\beta}}^{ols}$. The results obtained under the OLS are shown in the first two rows of Table \ref{tabla_prostate}, where $3/4$ of the total set has been used to fit the model (training set) and the remaining samples for the assessment of the generalization error of the resulting model (test set). The overall MSE, the prediction errors over the two groups as well as the number of coefficients involved in the model are presented. 

\begin{table*}[h!]
\begin{tabular}{cccccc}
\hline
\emph{Method} & & \emph{Overall MSE} & \emph{Group 1 MSE}& \emph{Group 2 MSE} & \emph{Non-Zero Coefficients}\\
\hline
\multirow{2}{*}{OLS}& Training set &  0.344 & 0.355 & 0.333 & \multirow{2}{*}{8}\\
 & Test set & 0.373 & 0.380 &  0.367 & \\
  \hline
  \hline
\multirow{2}{*}{Lasso}& Training set & 0.365 & 0.397 & 0.335 & \multirow{2}{*}{5}\\
& Test set & 0.408 & 0.414 & 0.403 & \\
\hline
\hline
\multirow{2}{*}{CSCLasso}& Training set  & 0.355 & 0.357 & 0.352 & \multirow{2}{*}{6}\\
 &Test set & 0.393 & 0.399 & 0.388 & \\
 \hline
\end{tabular}
\caption{Results obtained using {\tt prostate} dataset}
\label{tabla_prostate}
\end{table*}

Once the model is fitted, there are two fundamental criteria for evaluating its performance: the accuracy of prediction and the identification of significant predictors, which provides a good interpretation of the solution. It is well-known that $\boldsymbol{\hat{\beta}}^{ols}$ is not sparse, as can be observed from Table \ref{tabla_prostate} where the eight predictor variables have been used by the model. This fact has entailed new penalization techniques as ridge regression \cite{doi:10.1080/00401706.1970.10488634}, which is a continuous shrinkage method that achieves its better accuracy prediction through a bias-variance trade-off. Nevertheless, ridge regression is known not to be able to render a parsimonious solution. In contrast, best-subset selection (\cite{doi:10.2307/2985341}) achieves more sparse solutions, but it suffers from high variability and computational difficulties \cite{doi:10.1198/016214501753382273}. To overcome those shortcomings, \cite{10.2307/2346178} proposed the Lasso regularization technique, which achieves both estimation and selection of relevant predictors simultaneously by construction. Given $\mathbf{X}=[\mathbf{1}\mid\mathbf{x}_1\mid\ldots\mid\mathbf{x}_p]$ the predictor matrix; then, the Lasso solution can be defined as
\begin{equation}\label{lasso}
\boldsymbol{\hat{\beta}}^{Lasso}(\lambda)= \underset{\boldsymbol{\beta}}{\arg \min} \dfrac{1}{n}\|\mathbf{y} - \mathbf{X}\boldsymbol{\beta}\|^2+ \lambda \|(\beta_1,\ldots,\beta_p)\|_1
\end{equation}
where $\lambda\geq 0$ is a tuning parameter and $\|.\|_1$ is the $l_1$ norm. To visualize the effect of the penalty term in the Lasso formulation, consider the third and fourth rows in Table \ref{tabla_prostate}, which provide the results obtained under the Lasso for the {\tt prostate} dataset. In this case, in comparison with OLS results, a sparser, and therefore a more interpretable solution, has been obtained at the expense of slightly worsening MSE values.

One of the advantages of the Lasso is that the entire path of solutions can be found thanks to the LARS algorithm \cite{efron2004}. In addition, it is well-known that, under some conditions, the Lasso enjoys good theoretical and statistical properties (\cite{10.2307/2345967}, \cite{friedman2001elements}, \cite{buhlmann2011statistics}). However, the Lasso presents some limitations; in particular, the literature related to the Lasso has not undertaken the problem of fully controlling the accuracy prediction on certain individuals of interest. In the previous {\tt prostate} database, assume for instance that we are interested in fitting a sparse regression model to the dataset where, apart from obtaining a small overall mean squared error, also the prediction error for the \emph{young} individuals should not exceed a given threshold. In this paper we propose a Lasso-based model that allows for such aim, namely the cost-sensitive constrained Lasso, denoted from now on as CSCLasso. The results obtained for the {\tt prostate} database under the CSCLasso, whose definition and main properties shall be discussed in Section \ref{Newmethod} and \ref{TheoreticalResults}, are shown in the last two rows of Table \ref{tabla_prostate}. A threshold for the mean squared error over \emph{Group 1} is set equal to $0.357$, which represents an improvement of $10\%$ over the prediction error of the Lasso ($0.397$). Note that in the training set the MSE satisfies the imposed constraint, as expected. Also note that the improvement in \emph{Group 1} is at the expense of slightly increasing the prediction error over \emph{Group 2}. In terms of sparsity, the CSCLasso model has needed an additional predictor variable comparing to Lasso in order to comply with the constraint.

As it will be seen in Section \ref{Newmethod}, the novel approach is set up by adding convex quadratic constraints to the Lasso formulation. Other approaches have considered constrained versions of the Lasso before, see for example (\cite{Gareth2019}, \cite{doi:10.1080/10618600.2018.1473777}, \cite{TORRESBARRAN20181921}, \cite{HU201513} and references therein). In such works, equality or/and inequality linear constraints are considered for imposing prior knowledge and structure onto the coefficient estimates. In our approach instead, quadratic convex constraints are formulated and thus, our approach and results generalize those previously obtained in the literature.

Not only constrained versions of the Lasso can be found in the literature. Indeed, many different variants have been proposed. For example, in \cite{doi:10.1198/016214506000000735} adaptive weights for penalizing different coefficients in the $\emph{l}_1$ penalty are included as a way for fitting sparser models under more general conditions. Moreover, in the presence of highly correlated predictor variables (as is usual in microarray studies) or when predictors are structurally grouped (e.g. dummy variables), the Lasso sometimes does not perform well and, as a consequence, the \emph{elastic net} \cite{elasticnet} and the \emph{group lasso} (\cite{thegrouplasso}, \cite{sparsegrouplasso_Simon}) were proposed. They combine $\emph{l}_2$ and $\emph{l}_1$ penalties to try to select (or remove) the correlated or structured predictor variables together. Other extension is to consider
\begin{equation}\label{lassoA}
\boldsymbol{\hat{\beta}}^{Lasso}(\lambda)= \underset{\boldsymbol{\beta}}{\arg \min} \dfrac{1}{n}\|\mathbf{y} - \mathbf{X}\boldsymbol{\beta}\|^2+ \lambda \|\mathcal{A}\boldsymbol{\beta}\|_1.
\end{equation}
instead of \eqref{lasso}, where $\mathcal{A}$ is a fixed matrix (see \cite{generalizedLasso}). If $\mathcal{A}=\left(0|I^{p \times p}\right)$, then the Lasso objective function is obtained; however, other forms of $\mathcal{A}$ different from the identity can be found in the literature, see for example \cite{doi:10.1093/biomet/asw065}. In fact, various choices of $\mathcal{A}$ in \eqref{lassoA} define problems that are already well-known in the literature as the \emph{fused lasso} \cite{fusedlasso}. See \cite{StatisticalLearning} for an extensive review about Lasso problem and generalizations.

The motivation of this paper (controlling the performance measure on certain groups of interest) is not novel in the Data Analysis literature, and indeed, it has been exploited in classification contexts with the term  \emph{cost-sensitive learning} (\cite{Prati2015} and \cite{he2013imbalanced}). Many realworld problems, such as those found in medical diagnosis or credit card fraud detection, have asymmetric misclassification costs associated, since the consequences of wrong predictions across the classes may be very different. Therefore, for these problems, it is more important to achieve better classification rates for the individuals of interest (ill people, defaulting customers). See \cite{Carrizosa2008}, \cite{Sun}, \cite{Datta}, \cite{LEE201792}, \cite{Bradfor1998} and \cite{Freitas207} for more details and applications. Those methods are
based on adapting the classifier construction or adding parameters, among others. As some examples, consider \cite{Datta}, \cite{Carrizosa2008} and \cite{LEE201792}, which adapt the support vector machine (SVM) classifier. In \cite{Datta} the decision boundary shift is combined with unequal misclassification penalties. A biobjective problem is introduced in \cite{Carrizosa2008}, which simultaneous minimizes the misclassification rates. In \cite{LEE201792}, the authors propose a new weight adjustment factor that is applied to a weighted SVM.

This paper is structured as follows. In Section \ref{Newmethod}, the cost-sensitive constrained Lasso (CSCLasso) is introduced and some key issues are discussed. Section \ref{TheoreticalResults} considers theoretical properties of the CSCLasso, as the existence and uniqueness of solution, limit behaviour (in terms of the penalty parameter) and consistency. Section \ref{SimulationStudy} presents a detailed numerical analysis with both simulated and real datasets, and finally, some conclusions and extensions are provided in Section \ref{Conclusions}. Technical proofs are relegated to the Appendix.

\section{The cost-sensitive constrained Lasso: definition and key aspects}
\label{Newmethod}
This section presents the cost-sensitive constrained Lasso, which, as will be seen, is defined through an optimization problem with constraints related to prediction errors for individuals of interest. In addition, some computational details, as well as different key aspects concerning the tuning parameters of our proposal, are presented.

\subsection{Definition}\label{Definition}

The proposed CSCLasso is a novel variant of the Lasso where we shall demand that the prediction errors for the groups of interest are below certain threshold values,
\begin{align}
\begin{split}
\underset{\boldsymbol{\beta}}{\min} \hspace{0.5cm} & \dfrac{1}{n_0}\|\mathbf{y}_0-\mathbf{X}_0\boldsymbol{\beta}\|^{2} + \lambda \|\mathcal{A}\boldsymbol{\beta}\|_1 \\
 \mbox{s.t.} \hspace{0.5cm} &  \dfrac{1}{n_1}\| \mathbf{y}_1-\mathbf{X}_1\boldsymbol{\beta}\|^{2} -f_1 \leq 0,\\
& \,\,\,\,\,\,\,\,\,\,\,\,\,\,\,\,\,\,\,\,\,\,\,\,\,\,\,\,\, \vdots\\
 & \dfrac{1}{n_L}\| \mathbf{y}_L-\mathbf{X}_L\boldsymbol{\beta}\|^{2} -f_L \leq 0.
\label{CSCLasso}\end{split}
\end{align}
In the previous formulation, $(\mathbf{y}_0, \mathbf{X}_0)$ is the set of observations used to build the sparse model with overall minimum MSE, which can be the complete dataset $(\mathbf{y}, \mathbf{X})$, or a subset of smaller size. Additionally, let $(\mathbf{y}_l, \mathbf{X}_l)$, $l=1,\ldots,L$, define groups of interest (not necessarily disjoint), where the MSE predictions are to be controlled. Then, $n_l$ is the number of instances related to group $l$. Finally, $f=(f_1,\ldots,f_L)$ contains the different threshold values for the MSE on the different groups. The solution of optimization problem (\ref{CSCLasso}) will be denoted by $\boldsymbol{\hat{\beta}}^{CSCLasso}(\lambda)$. From the formulation (\ref{CSCLasso}) it is natural to wonder whether running a Lasso on just the groups of interest is more advantageous. However, if a single Lasso is run on the groups of interest, dramatically bad predictions can be obtained when the resulting model is applied to new observations outside those groups, which is not the case for our approach. The same issue arises when a different Lasso model is built on each group of interest, but, in addition, new observations are not given with their group of origin. Contrary to what happens with our novel approach (\ref{CSCLasso}), the $L$ predictions obtained through the $L$ different estimated Lasso models may not be suitable to give a final prediction for such new samples.

The proposed method can be formulated as a Lasso with weighted quadratic penalties in the objective function associated with the different groups, but finding real meaning to their parameters (one per group) to be chosen is not an easy task (see \cite{doi:10.1287/opre.49.1.169.11190} and the references therein) and the full control over the accuracy prediction on certain individuals of interest would disappear. However, the parameters $f=(f_1,\ldots,f_L)$ involved in our model have a clear interpretation and, in addition, this formulation enables us to bound the prediction errors in the different groups of interest. 

As an example, in the illustration of the method in Section \ref{introduction} related to the {\tt prostate} dataset, whereas the training set was used in the objective function with $\mathcal{A}=\left(0|I^{8 \times 8}\right)$, the prediction error over the \emph{young} population of the training set, $(\mathbf{y}_1, \mathbf{X}_1)$, is controlled through a performance constraint ($f_1=0.357$). In a real application, once the $L$ groups of interest are selected by the user, threshold values $f_1, \ldots, f_L$ have to be fixed. Note that these thresholds will depend directly on the dataset in question and the considered groups of interest. As a first option, they could be fixed by the user according to her demand, but therefore unfeasibility problems may appear when solving the CSCLasso problem (\ref{CSCLasso}). For that reason, in Section \ref{thechoice} two procedures for determining such threshold values so that (\ref{CSCLasso}) is feasible are given.

Next, some other aspects related to the formulation of the CSCLasso and its resolution will be discussed.

\subsection{Computational details}\label{ComputationalDetails}
The CSCLasso problem as defined by (\ref{CSCLasso}) is a non-differentiable convex optimization problem with quadratic and convex constraints. However, if we rewrite the non-differentiable term in \eqref{CSCLasso} as
\begin{equation*}
\mathcal{A}\boldsymbol{\beta}=\mathbf{u}^+ - \mathbf{u}^-,
\end{equation*}
where $\mathbf{u}^+=(u_1^+, \ldots, u_p^+)$ and $\mathbf{u}^-=(u_1^-, \ldots, u_p^-)$ are new vectors of positive auxiliary variables, a differentiable version for the CSCLasso problem (\ref{CSCLasso}) is obtained in a straightforward manner as

\begin{align*}
\begin{split}
\underset{\boldsymbol{\beta},\mathbf{u}^+,\mathbf{u}^-}{\min} \hspace{0.5cm}& \dfrac{1}{n_0}\| \mathbf{y}_0-\mathbf{X}_0\boldsymbol{\beta}\|^{2} + \lambda \sum_{j=1}^{p} u_j^+ + \lambda \sum_{j=1}^{p} u_j^-\\
\mbox{s.t.}\hspace{0.5cm} & \dfrac{1}{n_{1}}\|  \mathbf{y}_1-\mathbf{X}_1\boldsymbol{\beta}\|^{2} -f_1 \leq 0, \\
& \,\,\,\,\,\,\,\,\,\,\,\,\,\,\,\,\,\,\,\,\,\,\,\,\,\,\,\,\,\vdots\\
& \dfrac{1}{n_{L}}\| \mathbf{y}_L-\mathbf{X}_L\boldsymbol{\beta}\|^{2} -f_L \leq 0,\\
& \mathcal{A}\boldsymbol{\beta}=\mathbf{u}^+ - \mathbf{u}^-,\\
& \mathbf{u}^+, \mathbf{u}^- \geq 0.
\label{CSCLasso_diferentiable}\end{split}
\end{align*}

This previous smooth formulation for the CSCLasso eases its resolution notably, since efficient solvers for quadratically constrained programming problems, such as Gurobi \cite{gurobi}, are available. In particular, the Gurobi R interface will be used in this work to obtain all numerical results.

Another remark concerning the formulation of the CSCLasso is that, instead of using the sum of squared deviations, least absolute deviations could have been considered. Then, \eqref{CSCLasso} would be reduced to a regression problem under linear inequality constraints, as those described in \cite{Gareth2019}, \cite{doi:10.1080/10618600.2018.1473777} and \cite{HU201513}. Nevertheless, to cope the non-differentiability of the absolute value function, a huge number of constraints and new auxiliary variables, which would depend on $n$, should have been added. Consequently, these constrained approaches are likely to face severe numerical difficulties in practice for large datasets.

\subsection{The choice of threshold values}\label{thechoice}
As commented in Section \ref{Definition}, threshold values $f_1, \ldots, f_L$ could be fixed by the user. If the user is too demanding, imposing very low MSE threshold values for (some of) the different groups, the optimization problem may become unfeasible. Although a try-and-error procedure may be used, it would be very helpful to have strategies yielding feasible solutions. Here we propose two procedures for determining $f_1, \ldots, f_L$ in such a way that (\ref{CSCLasso}) is feasible.

First, we propose a choice of the threshold values so that they are close to the OLS results, 
\begin{equation}\label{fl}
f_l=(1+\tau)MSE_l(\boldsymbol{\hat{\beta}}^{ols}), \,\, l=1,\ldots,L,  \rlap{\footnotesize\quad \quad \quad \quad \quad \quad  \quad \quad \quad \quad \quad \quad \quad \quad  \,\,}
\end{equation}
where $MSE_l(\boldsymbol{\beta})= \dfrac{1}{n_l}\|\mathbf{y}_l - \mathbf{X}_l\boldsymbol{\beta}\|^2$,  $l=1,\ldots,L$ and $\tau \geq 0$ is a small parameter whose meaning is the percentage of worsening with respect to the OLS prediction error. For the numerical example in Section \ref{introduction}, we could have imposed the threshold for the MSE over \emph{Group 1} equal to $0.352$, which is a $10\%$ ($\tau=0.1$) more than $MSE_1(\boldsymbol{\hat{\beta}}^{ols})=0.319$. The choice \eqref{fl} deals with the heterogeneity coming from the variability of the different groups ($MSE_l$ is different across groups). Nevertheless, when heterogeneity related to the importance of each group is also considered, the parameter $\tau$ can be replaced in \eqref{fl} by $\tau_l$, $l=1,\ldots,L$.

Next, we shall compute the minimum value of $\tau$, $\tau_{min}$, so as to (\ref{CSCLasso}) is feasible. That is, the minimum $\tau$ so that there exists $\boldsymbol{\beta}^{*}$ satisfying

$$\left(\underset{l}{\max} \dfrac{MSE_l(\boldsymbol{\beta}^{*})}{MSE_l(\boldsymbol{\hat{\beta}}^{ols})}\right) - 1 \leq \tau,$$
and, therefore, $\tau_{min}$ will be given as
\begin{equation*}\label{tau_min}
\tau_{min}=\left(\underset{l}{\max} \dfrac{MSE_l(\boldsymbol{\beta}^{*})}{MSE_l(\boldsymbol{\hat{\beta}}^{ols})} \right)-1.
\end{equation*}

Such $\tau_{min}$ can be found as the optimal value of the following linear problem with convex and quadratic constraints

\begin{align}
\begin{split}
\underset{\boldsymbol{\beta},z}{\min} \hspace{0.5cm} & z\\
\mbox{s.t.} \hspace{0.5cm} & z\geq  \dfrac{MSE_l(\boldsymbol{\beta})}{MSE_l(\boldsymbol{\hat{\beta}}^{ols})} - 1, \,\, \forall l=1, \ldots,L.
\label{tau_min}\end{split}
\end{align}

The feasible version of the CSCLasso optimization problem can be formulated as

\begin{align}
\begin{split}
\underset{\boldsymbol{\beta}}{\min} \hspace{0.5cm}  &  \dfrac{1}{n_0}\|\mathbf{y}_0-\mathbf{X}_0\boldsymbol{\beta}\|^{2} + \lambda \|\mathcal{A}\boldsymbol{\beta}\|_1 \\
\mbox{s.t.} \hspace{0.5cm} & \dfrac{1}{n_1}\| \mathbf{y}_1-\mathbf{X}_1\boldsymbol{\beta}\|^{2} - (1+ \tau)MSE_1(\boldsymbol{\hat{\beta}}^{ols}) \leq 0,\\
& \,\,\,\,\,\,\,\,\,\,\,\,\,\,\,\,\,\,\,\,\,\,\,\,\,\,\,\,\,\vdots\\
& \dfrac{1}{n_L}\| \mathbf{y}_L-\mathbf{X}_L\boldsymbol{\beta}\|^{2} - (1+ \tau)MSE_L(\boldsymbol{\hat{\beta}}^{ols}) \leq 0,
\label{CSCLasso_feasible1}\end{split}
\end{align}
where $\tau \geq \tau_{min}$.

Finally, note that if $\tau$ is big enough, then solving (\ref{CSCLasso_feasible1}) is equivalent to solve the unconstrained problem. Indeed, it is possible to find the value of $\tau$, $\tau_{max}(\lambda)$, such that both the constrained and unconstrained problems are equivalent

\begin{equation}\label{tau_max_lambda}
\begin{aligned}
\tau_{max}(\lambda)= & \,\, \underset{l\in \{1,\ldots,L\}}{max}
& & \dfrac{MSE_l(\boldsymbol{\hat{\beta}}^{Lasso}(\lambda))}{MSE_l(\boldsymbol{\hat{\beta}}^{ols})} - 1.
\end{aligned}
\end{equation}

A second possible choice for the threshold values follows an analogous approach but, instead of considering the results of the OLS, we shall consider the mean squared error of the Lasso, as in the numerical example introduced in Section \ref{introduction}. For each $l=1,\ldots,L$, 
\begin{equation}\label{f0_}
f_l=(1-\gamma)MSE_l(\boldsymbol{\hat{\beta}}^{Lasso}(\lambda)), \,\, l=1,\ldots,L,  \rlap{\footnotesize\quad \quad \quad \quad \quad \quad  \quad \quad \quad \quad \, \,\,\,}
\end{equation}
where $\gamma \geq 0$ is related to the desired percentage of improvement over the Lasso solution ($\gamma=0.1$ in the numerical example of Section \ref{introduction}).
In this case, we will compute the maximum value of $\gamma$, $\gamma_{max}$, in such a way that (\ref{CSCLasso}) is feasible under \eqref{f0_}, and the linear problem associated with $\gamma_{max}$ is

\begin{align}
\begin{split}
\underset{\boldsymbol{\beta},z}{\max} \hspace{0.5cm} & z\\
\mbox{s.t.} \hspace{0.5cm} & 1-\dfrac{MSE_l(\boldsymbol{\beta})}{MSE_l(\boldsymbol{\hat{\beta}}^{Lasso}(\lambda))}\geq z, \,\, \forall l=1, \ldots,L.
\label{gamma_max}\end{split}
\end{align}
Thus, another possible feasible version of the CSCLasso optimization problem can be formulated as
\begin{align}
\begin{split}
\underset{\boldsymbol{\beta}}{\min} \hspace{0.5cm}  &  \dfrac{1}{n_0}\|\mathbf{y}_0-\mathbf{X}_0\boldsymbol{\beta}\|^{2} + \lambda \|\mathcal{A}\boldsymbol{\beta}\|_1 \\
\mbox{s.t.} \hspace{0.5cm} & \dfrac{1}{n_1}\| \mathbf{y}_1-\mathbf{X}_1\boldsymbol{\beta}\|^{2} - (1- \gamma)MSE_1(\boldsymbol{\hat{\beta}}^{Lasso}(\lambda)) \leq 0,\\
& \,\,\,\,\,\,\,\,\,\,\,\,\,\,\,\,\,\,\,\,\,\,\,\,\,\,\,\,\,\vdots\\
& \dfrac{1}{n_L}\| \mathbf{y}_L-\mathbf{X}_L\boldsymbol{\beta}\|^{2} - (1- \gamma)MSE_L(\boldsymbol{\hat{\beta}}^{Lasso}(\lambda)) \leq 0,
\label{CSCLasso_feasible2}\end{split}
\end{align}
where $\gamma \leq \gamma_{max}$.

Note that the two choices previously described for selecting the threshold values are not unique. Indeed, instead of using the MSE, another statistical measure as the R-squared can be considered. Further details about how they perform in numerical applications are described in Sections \ref{role} and \ref{SimulationStudy}.

\subsection{The role of the tuning parameters}\label{role}
The CSCLasso, as defined by (\ref{CSCLasso_feasible1}) or (\ref{CSCLasso_feasible2}), is stated in terms of two tuning parameters, $\lambda$ and $\tau$ or $\lambda$ and $\gamma$, respectively. The first one, $\lambda$, is related to the sparsity of the solution, and the second one is linked to the  user's demanding level, since the degree of requirement increases as $\tau\rightarrow \tau_{min}$ (or $\gamma\rightarrow \gamma_{max}$). In this section we investigate how the solution of the CSCLasso changes when $\lambda$ and $\tau$ jointly vary (analogous results are obtained if $\lambda$ and $\gamma$ are analyzed instead). With this purpose, consider again the experimental setting as in the example of Section \ref{introduction} related to {\tt prostate} dataset with $\mathcal{A}=\left(0|I^{p \times p}\right)$ (Lasso objective function), but in this occasion assume that the prediction errors of both groups (the \emph{young} and the \emph{elderly} people) shall be controlled.
\par
\begin{figure}[h!]
\centering
\subfigure{\includegraphics[width=120mm]{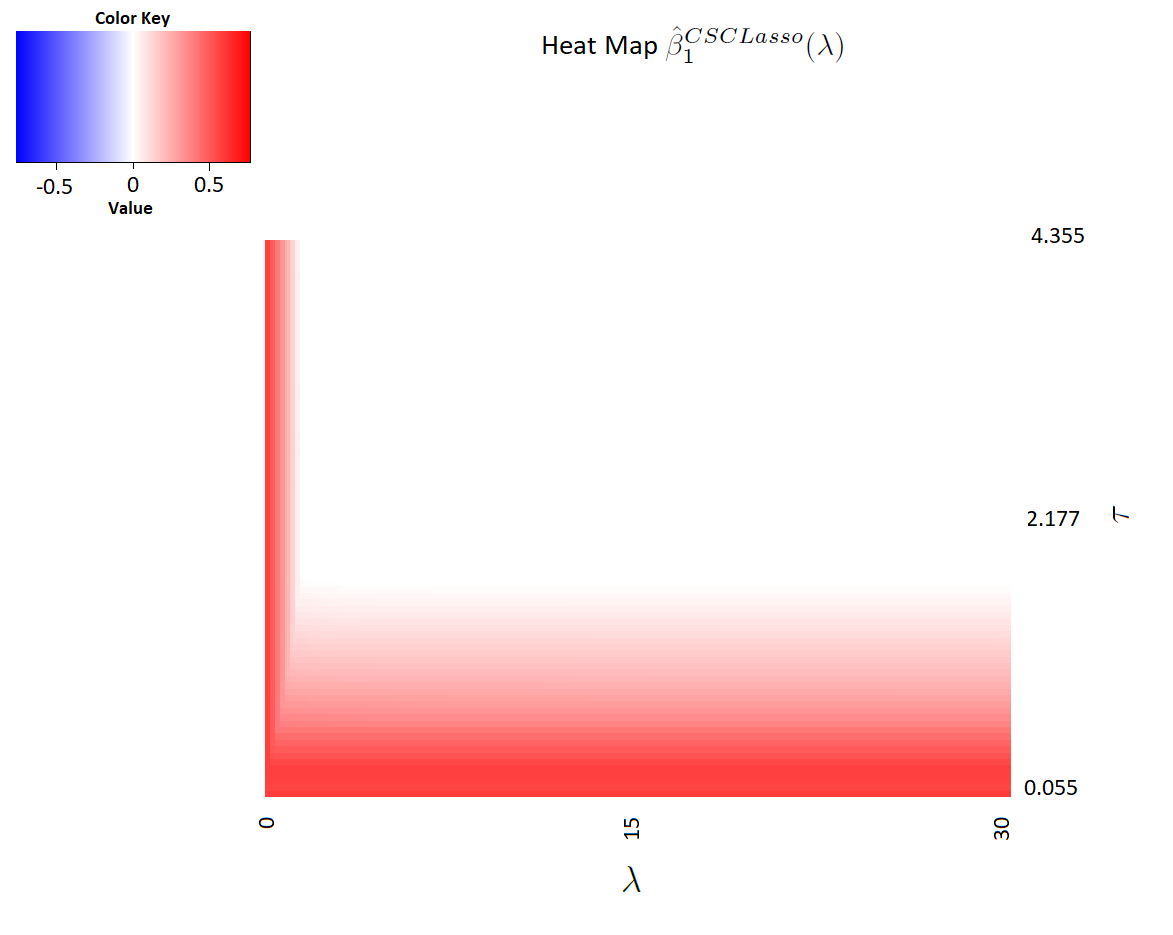}}
\caption{Heat map of $\hat{\beta}^{CSCLasso}_1(\lambda)$ using {\tt prostate} dataset}
\label{fig:heat_map_beta1}
\end{figure}

The interval of variation of the parameter $\lambda$ is set to $I_\lambda = [0,30]$. Moreover, according to (\ref{tau_min}), the smallest value of $\tau$ such that the CSCLasso optimization problem (\ref{CSCLasso_feasible1}) is feasible is $\tau_{min}=0.055$. On the other hand, following (\ref{tau_max_lambda}), $\tau_{max}=\underset{\lambda \in [0,30] }{\max} \,\,\tau_{max}(\lambda)=2.355$, although we will enlarge the interval of variation of $\tau$ to also visualize the unconstrained solution; such interval will be finally set as $I_\tau =[\tau_{min}, \tau_{max}+2]=[0.055, 4.355]$. Figure \ref{fig:heat_map_beta1} represents, via a heat map, the solution for $\hat{\beta}^{CSCLasso}_1(\lambda)$ for the different values of $(\lambda, \tau)$ in a grid contained in $I_\lambda \times I_\tau$.

Some conclusions can be drawn from the figure. Consider first the cases  where $\tau$ and $\lambda$ are big enough. Since, in this case, $\tau\geq\tau_{max}$, then, as commented at the end of the previous section, solving (\ref{CSCLasso_feasible1}) is equivalent to solving the Lasso. Therefore, $\boldsymbol{\hat{\beta}}^{CSCLasso}(\lambda)=\boldsymbol{\hat{\beta}}^{Lasso}(\lambda)=\boldsymbol{0}$ will be the optimal solution, provided that $\lambda$ is big enough. Analogously, if $\tau\geq\tau_{max}$ but $\lambda$ is small, then $\boldsymbol{\hat{\beta}}^{CSCLasso}(\lambda)=\boldsymbol{\hat{\beta}}^{Lasso}(\lambda)$, which will be equal to zero or not depending on the importance of the variable. When $\tau$ is small, the constraints are demanding and, even for large values of $\lambda$, it might happen that $\hat{\beta}^{CSCLasso}_1(\lambda)\neq \hat{\beta}^{Lasso}_1(\lambda)=0$, as it is the case.

Figure \ref{fig:Summarizing graphic} (see Appendix for further results) represents the analogous heat maps concerning $\hat{\beta}^{CSCLasso}_2(\lambda), \ldots, \hat{\beta}^{CSCLasso}_8(\lambda)$. A similar discussion as with $\hat{\beta}^{CSCLasso}_1(\lambda)$
is applicable to these figures. An interesting remark to be made concerns the importance of each variable: while variable $1$ is the only one selected for the Lasso, the CSCLasso returns a less sparse solution in this case, since predictor variables 1, 2, 4 and 5 turn out to be significant. However, this is not the rule, since there are examples where the level of sparsity is higher for the CSCLasso, as will be shown in Section \ref{SimulationStudy}.


\section{Theoretical properties}\label{TheoreticalResults}

In this section we discuss some theoretical results concerning the CSCLasso model. In Section \ref{ExandUniq}, the existence of a unique optimal solution to Problem (\ref{CSCLasso}) is proven for a fixed value of $\lambda \geq 0$. Section \ref{LimitBehaviour_lambda} deals with the limit behavior of the solution when $\lambda$ approaches infinity. Finally, some consistency properties of the CSCLasso solution are derived in Section \ref{Consistency} from the \emph{Sample Average Approximation} theory (see \cite{shapiro2009lectures}).

\subsection{Existence and uniqueness of solution}\label{ExandUniq}

For constrained versions of the Lasso in the literature, as in \cite{Gareth2019}, it is not possible to obtain the path of solutions and therefore approximations are made with the use of numerical algorithms. For the CSCLasso problem, a closed form solution of expression $\boldsymbol{\hat{\beta}}^{CSCLasso}(\lambda)$ is not available. However, an implicit characterization of the CSCLasso solution (with only one constraint) can be found as the following result states.

\begin{proposition}\label{Prop_1}
Consider the CSCLasso problem with one constraint,
\begin{align}
\begin{split}
\underset{\boldsymbol{\beta}}{\min} \hspace{0.5cm}  &  \dfrac{1}{n_0}\|\mathbf{y}_0-\mathbf{X}_0\boldsymbol{\beta}\|^{2} + \lambda \|\mathcal{A}\boldsymbol{\beta}\|_1 \\
\mbox{s.t.} \hspace{0.5cm} & \dfrac{1}{n_1}\| \mathbf{y}_1-\mathbf{X}_1\boldsymbol{\beta}\|^{2} - (1+ \tau)MSE_1(\boldsymbol{\hat{\beta}}^{ols}) \leq 0,\\
\label{CSCLassoOneConst}\end{split}
\end{align}
where $\mathcal{A}=\left(0|I^{p \times p}\right)$ and assume that $\mathbf{X}_0$ and $\mathbf{X}_1$ are maximum rank matrices. Then 
\begin{equation}\label{BetaCSCLassosimple}
\boldsymbol{\hat{\beta}}^{CSCLasso}(\lambda)=\left(\dfrac{1}{n_0}\mathbf{X}_0'\mathbf{X}_0 + \dfrac{1}{n_1}\eta(\lambda)\mathbf{X}_1'\mathbf{X}_1\right)^{-1}\left(\dfrac{1}{n_0}\mathbf{X}_0'\mathbf{y}_0+\dfrac{1}{n_1}\eta(\lambda)\mathbf{X}_1'\mathbf{y}_1\right)-\dfrac{1}{2}\left(\dfrac{1}{n_0}\mathbf{X}_0'\mathbf{X}_0 + \dfrac{1}{n_1}\eta(\lambda)\mathbf{X}_1'\mathbf{X}_1\right)^{-1}\mathbf{b}(\lambda)
\end{equation}
where $\eta(\lambda)$ is the Lagrange multiplier of the constraint and the component $s$, $s=0,1,\ldots,p$, of the vector $\mathbf{b}(\lambda)$ is given by
$$b_s(\lambda)=\left\{ \begin{array}{lcc}
              \,\,\,\,\, \lambda,  \,\, \text{if}  \,\,  \hat{\beta}^{CSCLasso}_s(\lambda)>0, \\
             -\lambda,  \,\,  \text{if} \,\,  \hat{\beta}^{CSCLasso}_s(\lambda)<0, \\
             \,\,\,\,\, 0  \,\,  \,\,  else. \\
             \end{array}\right.$$\\
\end{proposition}

From the previous proposition, it is clear that a closed form solution is hard to be obtained, even in the simplest scenario.
 
\begin{figure}[h!]
\centering
\subfigure{\includegraphics[width=112mm]{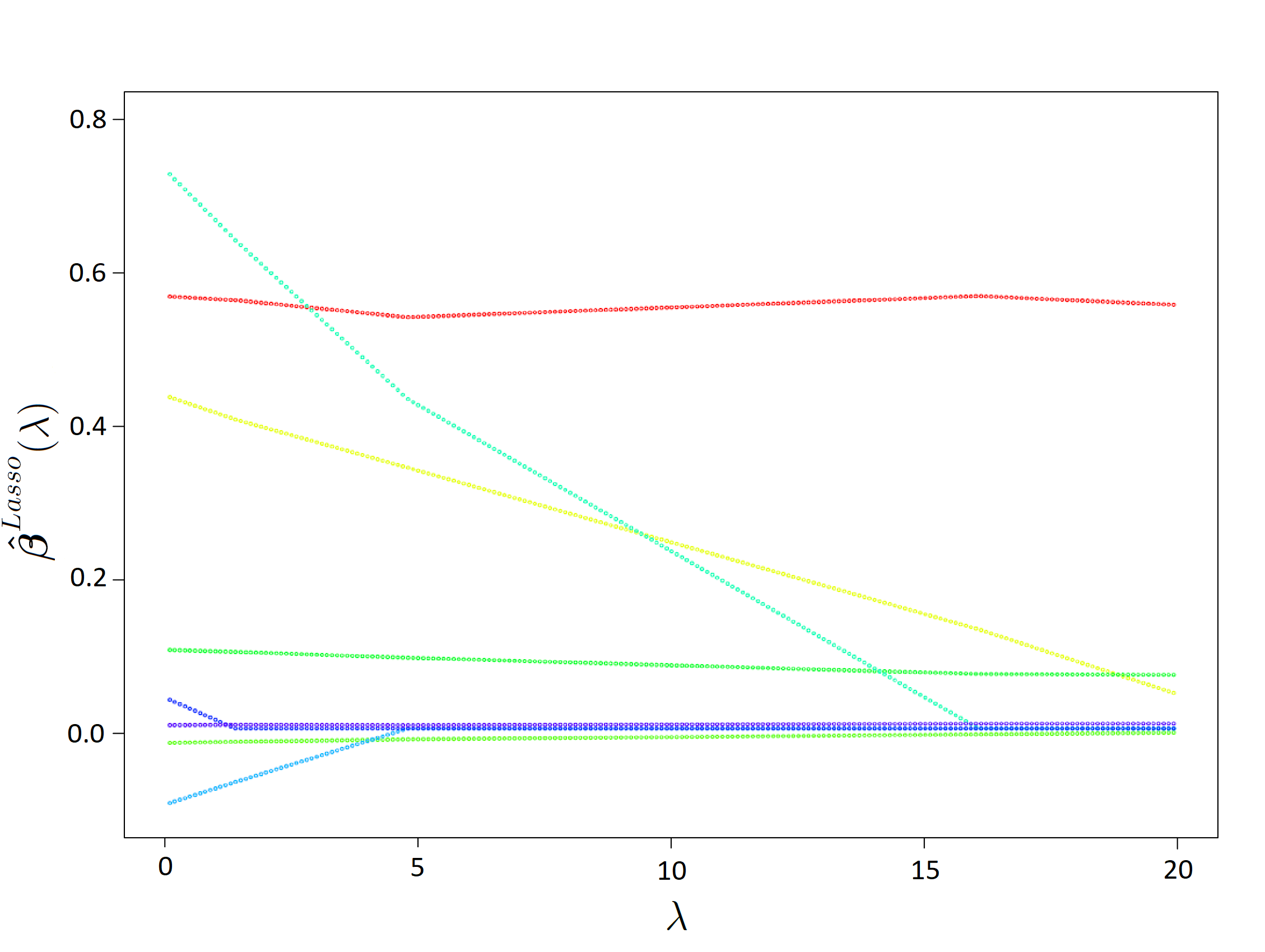}}
\subfigure{\includegraphics[width=112mm]{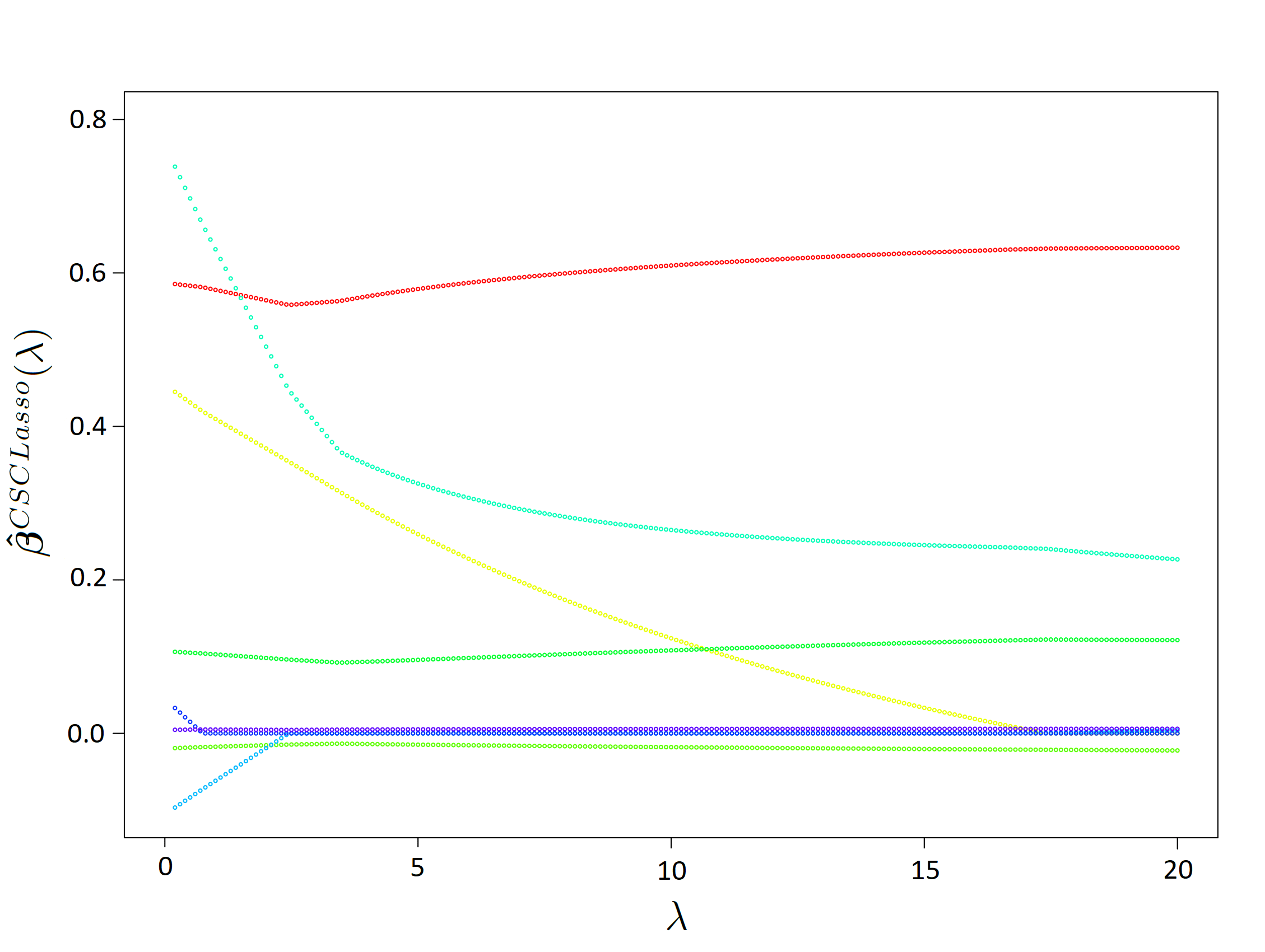}}
\caption{Path of solutions under Lasso (top) and CSCLasso (bottom) for {\tt prostate} dataset }
 \label{fig:CSCLasso_oneconstraint}
\end{figure}

Nevertheless, given a fixed value of $\lambda$, the CSCLasso problem can be solved using quadratic programming via any of the standard solvers available in the literature. As an example, Figure \ref{fig:CSCLasso_oneconstraint} depicts the path of solutions for the {\tt prostate} example introduced in Section \ref{introduction}, for an assortment of values of $\lambda$ in a grid (see Section \ref{LimitBehaviour_lambda} for details). Each line represents a component of $\boldsymbol{\hat{\beta}}^{CSCLasso}(\lambda)$, $\hat{\beta}^{CSCLasso}_j(\lambda)$ with $j=1,\ldots,8$. It can be observed from the figure that, contrary to what happens in the Lasso path of solutions (top panel of Figure \ref{fig:CSCLasso_oneconstraint}), the CSCLasso path of solutions is not piecewise linear (bottom panel of Figure \ref{fig:CSCLasso_oneconstraint}). Such non-linearity (due to the quadratic constraints) hinders the application of an iterative algorithm to obtain the path of solutions as those given in papers \cite{Gareth2019} and \cite{doi:10.1080/10618600.2018.1473777}.

\begin{figure}[h!]
\centering
\subfigure{\includegraphics[width=112mm]{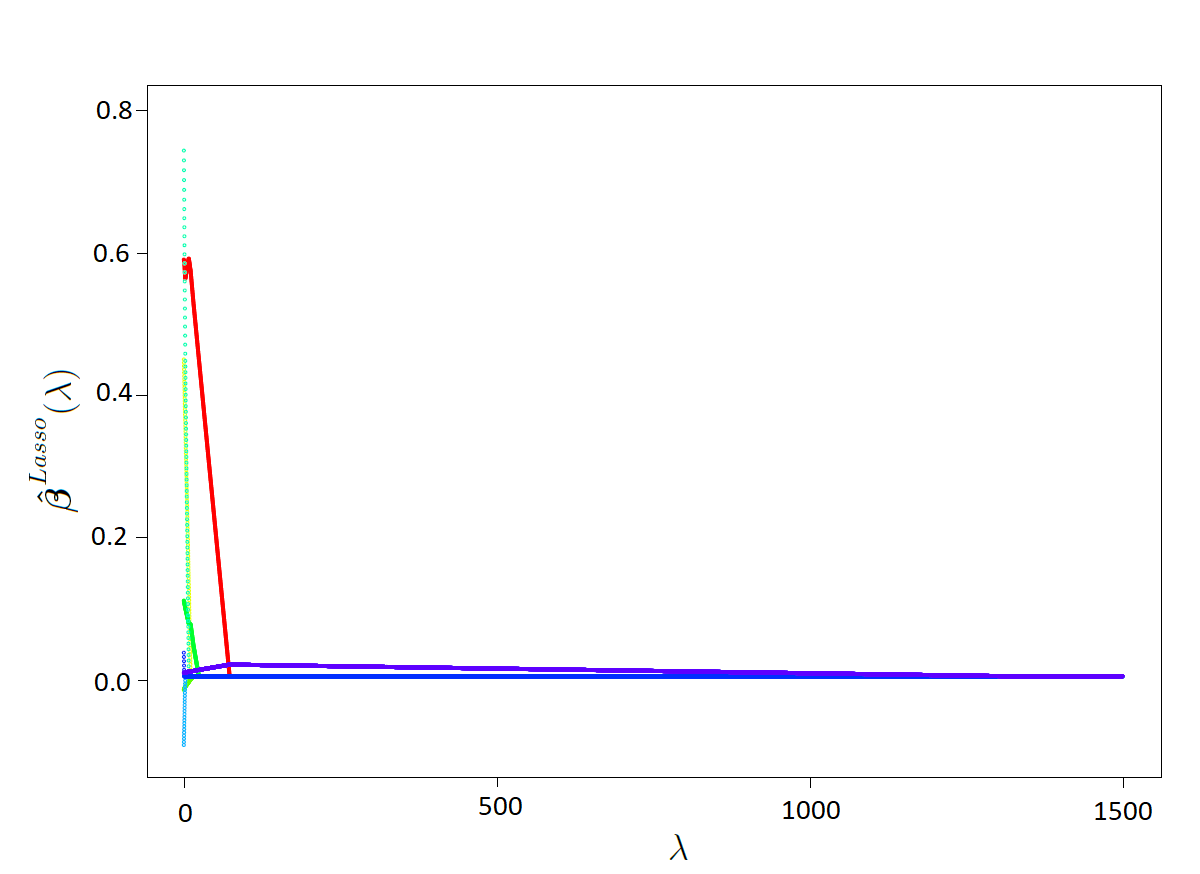}}
\subfigure{\includegraphics[width=112mm]{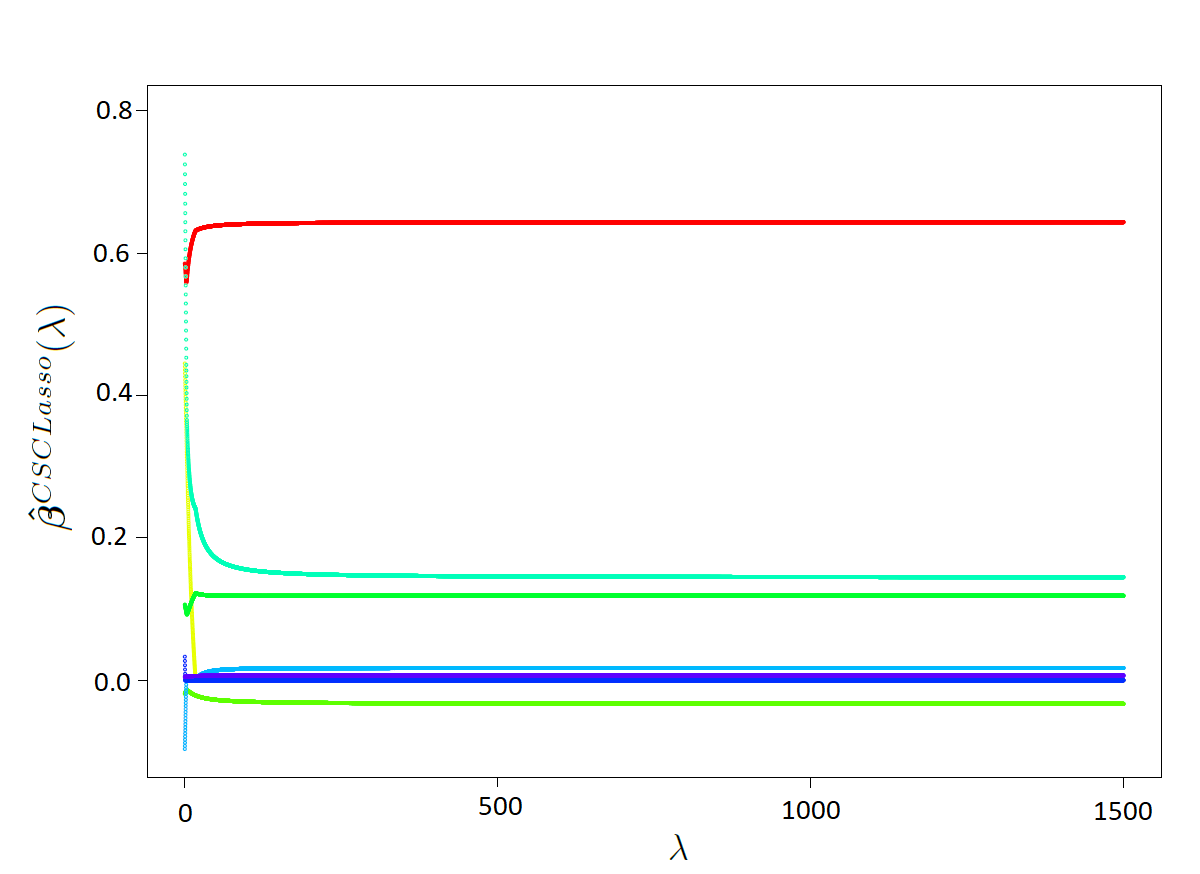}}
\caption{Path of solutions under Lasso (top) and CSCLasso (bottom) for {\tt prostate} dataset when $\lambda$ increases} \label{fig:CSCLasso_oneconstraint_1500}
\end{figure}

Also note from Figure \ref{fig:CSCLasso_oneconstraint_1500} that as a consequence of the performance constraints, the solution is stabilized when $\lambda$ increases, but does not shrink to 0, as with Lasso. This is detailed in Section \ref{LimitBehaviour_lambda}.

Even without having the expression of the general solution of (\ref{CSCLasso}), we next prove that, under full rank assumptions, the solution is unique. First, in order to simplify the formulation of (\ref{CSCLasso}), henceforth its feasible set will be denoted by $\boldsymbol{B}$, which is convex and closed. This is also true (and the results which follow remain valid) if, on top of the performance constraints, one adds linear constraints modeling other aspects of interest (for example, the sign of a certain coefficient can be fixed to be positive or negative depending on the known relation between the corresponding predictors with the response variable). In the same vein, henceforth, $(\mathbf{y}_0,\mathbf{X}_0) = (\mathbf{y},\mathbf{X})$ is used to minimize the overall MSE. In this way, the CSCLasso problem (\ref{CSCLasso}) is rewritten as
\begin{align}
\begin{split}
 \underset{\boldsymbol{\beta}\in \boldsymbol{B}}{\min} \hspace{0.5cm}& \dfrac{1}{n}\|\mathbf{y}-\mathbf{X}\boldsymbol{\beta}\|^{2} + \lambda \|\mathcal{A}\boldsymbol{\beta}\|_1.
\label{CSCLasso_GeneralB}\end{split}
\end{align}

The following result guarantees that the solution of problem (\ref{CSCLasso_GeneralB}) is unique. 
\begin{thm}\label{TheoExisandUniq}
Consider Problem (\ref{CSCLasso_GeneralB}) where $\mathbf{X}$ is assumed to be a maximum rank matrix and its feasible region $\boldsymbol{B}$ is a convex and closed set in $\mathbb{R}^{p+1}$. Then, Problem (\ref{CSCLasso_GeneralB}) has a unique optimal solution.
\end{thm}

\subsection{Asymptotic behaviour}\label{LimitBehaviour_lambda}  
One of the key points when dealing with Lasso-type problems is the choice of the regularization parameter $\lambda$. In the case of the Lasso, such choice is straightforward since the entire path of solutions is known to be piecewise linear, shrinking to 0. In  particular, it is known that there exists a value of $\lambda$, $\lambda^*$, such that the solution $\boldsymbol{\hat{\beta}}^{Lasso}(\lambda)=\mathbf{0}$ is optimal for all values $\lambda \geq \lambda^*$. 

The following result provides explicitly the value of such $\lambda^*$.

\begin{proposition}\label{Prop_2}
Consider the Lasso model (\ref{lassoA}). Define $\lambda^*$ as the optimal value of the linear programming problem
\begin{align*}
\begin{split}
\underset{z, \mathbf{t}}{\min}  \hspace{0.5cm}& z\\
 \mbox{s.t.}  \hspace{0.5cm}&  \dfrac{2}{n}\mathbf{X}'\mathbf{y} = \mathcal{A}'\lambda\mathbf{t},\\
 & -z \leq  \lambda t_s  \leq z, \quad s=0,1,\ldots,p.
\end{split}
\end{align*}
Then, $\boldsymbol{\hat{\beta}}^{Lasso}(\lambda)=\mathbf{0}$ for all $\lambda\geq \lambda^*$. In particular, for $\mathcal{A} = (0| I^{p\times p})$, 
$$\lambda^* = \left\|\dfrac{2}{n}\mathbf{X}'\mathbf{y} \right\|_{\infty}$$
\end{proposition}

Some works dealing with (linear) constrained versions of the Lasso (as \cite{doi:10.1080/10618600.2018.1473777} and \cite{Gareth2019}) have developed efficient algorithms to build the associated solutions path.
Consider now the general problem (\ref{CSCLasso_GeneralB}). As commented in Section \ref{ExandUniq}, the expression of $\boldsymbol{\hat{\beta}}^{CSCLasso}(\lambda)$ is not available in closed form and, consequently, the entire path cannot be computed. In this case, when $\lambda$ tends to $+\infty$, the solution $\boldsymbol{\hat{\beta}}^{CSCLasso}(\lambda)$ stabilizes around $\boldsymbol{\hat{\beta}}^{CSCLasso}(+\infty)=\underset{\boldsymbol{\beta}\in \boldsymbol{B}}{\arg \min} \| \mathcal{A}\boldsymbol{\beta}\|_1$. This idea is also used in \cite{doi:10.1080/10618600.2018.1473777} and \cite{Gareth2019}, where, in order to find an initialization for the algorithms, the proposed constrained problems are solved by only considering the penalty term in the objective function.
Such a limit solution is obtained by solving an optimization problem with a linear objective function and convex quadratic constraints, namely
\begin{align*}
\begin{split}
\underset{\boldsymbol{\beta}, \boldsymbol{u}^+, \boldsymbol{u}^-}{\min} \hspace{0.5cm}& \sum_{s=0}^{p} (u^+_s + u^-_s)\\
 \mbox{s.t.}  \hspace{0.5cm}&  \boldsymbol{\beta}\in \boldsymbol{B}\\
 & \mathcal{A}\boldsymbol{\beta}= \boldsymbol{u}^+ -\boldsymbol{u}^-\\
 & \boldsymbol{u}^+, \boldsymbol{u}^- \geq \boldsymbol{0}.
\end{split}
\end{align*}

A grid search is carried out in the general CSCLasso problem (\ref{CSCLasso_GeneralB}) to obtain suitable values of $\lambda$. In order to fix the grid, we propose the following dynamic approach to find an approximate maximum value of $\lambda$, $\lambda^*$ (see Algorithm \ref{tab:pseudo}). 

\begin{algorithm}[h]
\vspace*{0.2cm}
1. Fix $\varepsilon>0$ and $\boldsymbol{c}=(2^{-5})$. Fix $i=1$ and compute $\boldsymbol{\hat{\beta}}^{CSCLasso}(\boldsymbol{c}[i])$.\\
2. Compute $\boldsymbol{\hat{\beta}}^{CSCLasso}(+\infty)=\underset{\boldsymbol{\beta}\in \boldsymbol{B}}{\arg \min} \| \mathcal{A}\boldsymbol{\beta}\|_1$.\\
3. While $\| \boldsymbol{\hat{\beta}}^{CSCLasso}(\lambda) -  \boldsymbol{\hat{\beta}}^{CSCLasso}(+\infty)\| > \varepsilon$, repeat
\begin{itemize}
\item[a)] $i=i+1$
\item[b)] $\boldsymbol{c}=(\boldsymbol{c}, 2\boldsymbol{c}[i-1])$
\item[c)] compute $\boldsymbol{\hat{\beta}}^{CSCLasso}(\boldsymbol{c}[i])$
\end{itemize}
4. $\lambda^*=\boldsymbol{c}[i]$
\caption{Dynamic approach for selecting $\lambda^*$ in the CSCLasso}\label{tab:pseudo}
\vspace*{0.2cm}
\end{algorithm}

Once the maximum value $\lambda^*$ is found, then the grid ranges from $0$ to $\lambda^*$ with the desired step. Note that the previous algorithm already provides an initial grid of the form $(2^{-5}, 2^{-4},\ldots, 2^{0}, \ldots, \lambda^*)$.

\subsection{Consistency properties in the CSCLasso}
\label{Consistency}

The purpose of this section is to prove some results related to the consistency of both the solution and the objective value for CSCLasso problem (\ref{CSCLasso_GeneralB}). To do that, the theory of \emph{Sample Average Approximation} (SAA) \cite{shapiro2009lectures} will be applied. Consider the following stochastic programming problem
\begin{align}
\begin{split}
\min_{\boldsymbol{\beta} \in \boldsymbol{B}} \hspace{0.5cm} &  f(\boldsymbol{\beta}):=E[F(\boldsymbol{\beta},(Y,X))],
\label{trueproblem}\end{split}
\end{align}
where $\boldsymbol{B}$ is a nonempty closed subset of $\mathbb{R}^{p+1}$, $(Y,X)$ is an absolutely continuous random vector whose probability distribution $P$ is supported on a set $\Xi \subset \mathbb{R}^{p+1}$ and $F: \boldsymbol{B}\times \Xi \rightarrow \mathbb{R}$. In \cite{shapiro2009lectures}, under some conditions, the \emph{true} problem (\ref{trueproblem}) can be estimated by the SAA:
\begin{align}
\begin{split}
\min_{\boldsymbol{\beta} \in \boldsymbol{B}} \hspace{0.5cm} & \hat{f}_n(\boldsymbol{\beta}):=\dfrac{1}{n}\sum_{i=1}^{n}F(\boldsymbol{\beta},(y_i,x_i)),
\label{SAAproblem}\end{split}
\end{align}
where $x_i= (1, x_{i1}, \ldots, x_{ip})'$, and $\{(y_i,x_i)\}_{i=1,\ldots,n}$  is a realization of the $n$ random vectors $\{(Y_i,X_i)\}_{i=1,\ldots,n}$, which are independent and identically distributed (i.i.d.) as the random vector $(Y,X)$.  Note that the CSCLasso problem as in (\ref{CSCLasso_GeneralB}) takes the form of (\ref{SAAproblem}) as
\begin{equation}\label{PN}
\min_{\boldsymbol{\beta} \in \boldsymbol{B}} \dfrac{1}{n}\sum_{i=1}^{n}(y_{i}-x^{'}_{i} \boldsymbol{\beta})^{2}+ \lambda \| \mathcal{A}\boldsymbol{\beta}\|_1
\end{equation}

and the \emph{true} CSCLasso problem equivalent to (\ref{trueproblem}) is
\begin{equation}\label{P}
\min_{\mathbf{\boldsymbol{\beta}} \in \boldsymbol{B}} E[(Y-X'\mathbf{\boldsymbol{\beta}})^{2}+ \lambda \| \mathcal{A}\mathbf{\boldsymbol{\beta}}\|_1].
\end{equation}

Before proving the main result on the consistency of the CSCLasso, we first show the uniqueness of the solution of such a problem.

\begin{proposition}\label{uniqueness_teo} The optimal solution of the true CSCLasso problem (\ref{P}) is unique.
\end{proposition}

Denote by $\nu^{CSCLasso}(\lambda)$ and $\boldsymbol{\beta}^{CSCLasso}(\lambda)$, respectively,  the optimal value and the optimal solution of problem (\ref{P}). Analogously, let  $\hat{\nu}^{CSCLasso}(\lambda)$ and $\boldsymbol{\hat{\beta}}^{CSCLasso}(\lambda)$ be the optimal value and the optimal solution, respectively, of the SAA CSCLasso problem (\ref{PN}). The following result shows the consistency of the SAA values to the \emph{true} values.

\begin{thm}\label{consis_theo} Assume that $E[\|X\|^2] < \infty$, $E[Y^2]< \infty$, $E[\|YX\|]<\infty$. Then, $\hat{\nu}^{CSCLasso}(\lambda)$ converges to $\nu^{CSCLasso}(\lambda)$ and $\boldsymbol{\hat{\beta}}^{CSCLasso}(\lambda)$ converges to $\boldsymbol{\beta}^{CSCLasso}(\lambda)$ with probability one (w.p.\ 1).
\end{thm}

Finally, note that the theoretical results that have been studied in this work are also applicable to other versions of constrained Lasso as long as the feasible set is convex and closed, as is the case with the above-mentioned works \cite{Gareth2019}, \cite{doi:10.1080/10618600.2018.1473777} and \cite{HU201513}.

\section{Numerical experiments}\label{SimulationStudy}

In this section, the behaviour and performance of our approach is illustrated throughout an extensive empirical study. In particular, using both simulated and real datasets, the aim of the experiments shall be to improve the prediction errors of the Lasso in one or more groups of interest. Or in other words, threshold values shall be fixed as in (\ref{CSCLasso_feasible2}). Since our proposal is a novel extension of the Lasso, we will also show the results under the Lasso, not only for those groups that are controlled (for which obviously, the Lasso performs worse) but also for the non-controlled groups. In this way the CSCLasso can be better inspected in comparison to the Lasso. Other aspects as the overall MSE and the percentage of non-zero coefficients in the regression model, among others, will be explored. Such measures will be estimated through median values using a $5$-fold cross validation approach. To this end, the dataset will be split at each fold into three sets: the so-called training, validation and test sets. The training set is used to fit the model, the validation set is used to estimate prediction error for model selection and the test set is used for assessment of the generalization error of the final chosen model. 

\subsection{A simulation study}\label{simulateddatasets}
The generation of the synthetic datasets in this section follows that of \cite{doi:10.1093/biomet/asw065}, where an overparameterized regression model is considered to cope with stratified data. A number of groups $K=20$ is set and two different sample sizes per group are considered, $n_k=\{150,500\}$, for $k=1,\ldots,K$. The number of predictors $p$ will be chosen from $\{20, 100, 500\}$.
The matrix of predictor values $\mathbf{X}$ is generated according to a multivariate normal distribution with zero mean and covariance matrix $\Sigma$ being a Toeplitz matrix with element $(i,j)$ equal to $0.5^{\mid i-j\mid}$. Regarding the response vector, a set of 20 predictors are randomly selected (with indexes included in a set $P_0$), while the rest of predictors are noise (that is, $\beta_j=0$ for $j \notin P_0$). The coefficients of the significant 20 predictors are chosen as follows. First, consider 10 random predictors out of the 20 selected. For such predictors, if the group $k>6$ then $\beta_j=1$ and, otherwise, $\beta_j=1+K^{\frac{1}{2}}$. For the other 10 predictors, $\beta_j=1$ if $k\leq 6$ and $\beta_j=1+K^{\frac{1}{2}}$ otherwise. In this way, the predictors behave differently depending on the group. Finally, the response vector for each group is generated according to the standard linear regression model with normal error.

Once the synthetic dataset is built and its response and predictor variables have been standardized, the CSCLasso with $\mathcal{A}= (0|I^{p\times p})$ is run with constraints imposed over the first six groups. The choice of $\lambda$ will change at each fold. A grid in $\lambda$ is built as in Section \ref{LimitBehaviour_lambda}, and, the value of $\lambda$ which leads to the lower overall MSE in the validation set is selected. Table \ref{tabla_Simulados} shows the median prediction errors per group $k$ ($MSE_k$), $k=1,\ldots,6$, obtained by the Lasso (rows in grey color) and the corresponding values obtained under the CSCLasso for different threshold values $f$.
\begin{table*}[h!]
\tiny
\hspace*{-1.8cm}\begin{tabular}{ccccccccc|cccccc}
\cline{4-15}
  && &\multicolumn{6}{c}{$n_k=150$} & \multicolumn{6}{c}{$n_k=300$}  \\
  \hline
$p$ & \multicolumn{2}{c}{} & Group 1& Group 2 & Group 3& Group 4 & Group 5& Group 6  & Group 1& Group 2 & Group 3& Group 4 & Group 5& Group 6 \\
\hline
  \multirow{13}{*}{20} & \multicolumn{2}{c}{\cellcolor{lightgray}$MSE_k$(\emph{Lasso})}  & \cellcolor{lightgray}1.084 & \cellcolor{lightgray}0.904 & \cellcolor{lightgray}0.768 & \cellcolor{lightgray}0.925 &\cellcolor{lightgray}0.944  & \cellcolor{lightgray}0.674 & \cellcolor{lightgray}0.801& \cellcolor{lightgray}0.770& \cellcolor{lightgray}0.966& \cellcolor{lightgray}0.937& \cellcolor{lightgray}0.776& \cellcolor{lightgray}0.938\\
 \cline{2-15}
&  \emph{Improv.}  & $f$ &  1.052 & 0.877 & 0.745 & 0.898 & 0.916 & 0.653 &  0.777& 0.747& 0.937 & 0.909& 0.753 & 0.910\\
&3\% &$MSE_k$ & 0.933 & 0.722 & 0.695 & 0.859 & 0.822 &  0.574 &0.752& 0.689 &0.810 & 0.859 &0.744 & 0.870 \\
  \cline{2-15}
& \emph{Improv.}&  $f$ & 1.030  & 0.859  & 0.730  & 0.879 & 0.897 & 0.640 & 0.761 & 0.732& 0.918 & 0.890 & 0.737 &0.891  \\
 & 5\% &$MSE_k$ & 0.921 & 0.708 & 0.689 & 0.848 & 0.804 & 0.564 & 0.734 & 0.682& 0.787 & 0.844 & 0.736 & 0.859\\
  \cline{2-15}
 & \emph{Improv.} & &\multirow{2}{*}{-} & \multirow{2}{*}{-} &\multirow{2}{*}{-} &\multirow{2}{*}{-} & \multirow{2}{*}{-} & \multirow{2}{*}{-} & 0.745 & 0.717 & 0.899 & 0.871 & 0.722 & 0.873\\
& 7\% &\emph{} &  &  &  &  &  &    &0.717 & 0.677& 0.768& 0.834& 0.721  & 0.847\\
   \cline{2-15}
 &\emph{Improv.} & &\multirow{2}{*}{-}& \multirow{2}{*}{-} &\multirow{2}{*}{-} &\multirow{2}{*}{-} & \multirow{2}{*}{-} & \multirow{2}{*}{-} &\multirow{2}{*}{-} &\multirow{2}{*}{-} &\multirow{2}{*}{-} & \multirow{2}{*}{-}& \multirow{2}{*}{-}& \multirow{2}{*}{-}\\
   & 10\% &   & &  & &  &  &  & & & & \\
    \cline{2-15}
 & \emph{Improv.}& & \multirow{2}{*}{-} & \multirow{2}{*}{-}&\multirow{2}{*}{-} &\multirow{2}{*}{-} & \multirow{2}{*}{-}& \multirow{2}{*}{-}&\multirow{2}{*}{-} &\multirow{2}{*}{-} &\multirow{2}{*}{-} & \multirow{2}{*}{-}& \multirow{2}{*}{-}& \multirow{2}{*}{-}\\
   & 15\% &     & & &  &  & &  & & & & \\
  \cline{2-15}
 & \emph{Improv.} &  &\multirow{2}{*}{-} & \multirow{2}{*}{-}& \multirow{2}{*}{-} & \multirow{2}{*}{-}& \multirow{2}{*}{-}& \multirow{2}{*}{-} &\multirow{2}{*}{-} &\multirow{2}{*}{-} &\multirow{2}{*}{-} & \multirow{2}{*}{-}& \multirow{2}{*}{-}& \multirow{2}{*}{-}\\
   & 20\% &     & & &  &  & &  & & & & \\
 \hline
   \multirow{13}{*}{100}  & \multicolumn{2}{c}{\cellcolor{lightgray}$MSE_k$(\emph{Lasso})}  & \cellcolor{lightgray}1.491 & \cellcolor{lightgray}1.139 & \cellcolor{lightgray}0.875 & \cellcolor{lightgray}1.496 & \cellcolor{lightgray}1.151 &  \cellcolor{lightgray}0.759 &\cellcolor{lightgray}1.104& \cellcolor{lightgray}1.151 & \cellcolor{lightgray}0.932& \cellcolor{lightgray}0.981& \cellcolor{lightgray}1.311 & \cellcolor{lightgray}1.142  \\
    \cline{2-15}
& \emph{Improv.}  & $f$ & 1.447 & 1.105 & 0.848 & 1.451 & 1.116 & 0.736 & 1.070 & 1.117 & 0.904 & 0.951 & 1.272 & 1.108 \\
 & 3\% &$MSE_k$   & 1.234 & 0.948 & 0.781  & 1.322 & 0.945 & 0.708 &1.044 & 1.145 & 0.911 & 0.963 & 1.306& 1.139 \\
  \cline{2-15}
 & \emph{Improv.}  & $f$ & 1.417 & 1.082 & 0.831 & 1.421 & 1.093 & 0.721 & 1.048 & 1.094 & 0.885 & 0.932 & 1.245 & 1.085 \\
  & 5\% &$MSE_k$   & 1.226 & 0.941 & 0.777 & 1.317 & 0.937 & 0.706 & 1.025 & 1.123 & 0.897 & 0.947 & 1.311 & 1.142\\
   \cline{2-15}
 &  \emph{Improv.}  & $f$ & 1.387 & 1.059 & 0.813 & 1.391 & 1.070 & 0.706 & 1.026 & 1.071 & 0.866 & 0.912 & 1.219 & 1.062 \\
  & 7\% &$MSE_k$  & 1.219 & 0.933 & 0.773 & 1.311 & 0.929 & 0.704 & 1.005& 1.111 & 0.883& 0.934 & 1.309 & 1.141 \\
   \cline{2-15}
 &  \emph{Improv.} & $f$ & 1.342
  & 1.025 & 0.787 & 1.346 & 1.035 & 0.683 & 0.993 & 1.036 & 0.838 & 0.883 & 1.180 & 1.028 \\
   & 10\%  &$MSE_k$  & 1.207 & 0.921 & 0.767  & 1.303 & 0.917 &  0.702 & 0.978& 1.094& 0.864& 0.915 & 1.311 & 1.131 \\
    \cline{2-15}
&  \emph{Improv.} & $f$& 1.268  & 0.968 & 0.743 & 1.271 & 0.978 & 0.645 & 0.938 & 0.979 & 0.792 & 0.834 & 1.114 & 0.971  \\
  & 15\% &$MSE_k$  & 1.128 & 0.903 & 0.765 & 1.250 & 0.919 & 0.695& 0.936& 1.061 & 0.833& 0.882& 1.274 & 1.087 \\
   \cline{2-15}
 &  \emph{Improv.} & $f$  & 1.193 & 0.911 & 0.700 & 1.196 & 0.920 & 0.607 & 0.883 & 0.921  & 0.745 & 0.785 & 1.049 & 0.914 \\
   &20\% &$MSE_k$  & 1.155 & 0.909 & 0.771& 1.251 & 0.900& 0.700 & 0.895 & 1.036& 0.803& 0.855& 1.227 & 1.046 \\ \hline

     \multirow{13}{*}{500}  & \multicolumn{2}{c}{\cellcolor{lightgray}$MSE_k$(\emph{Lasso})} & \cellcolor{lightgray}1.306 & \cellcolor{lightgray}1.133 & \cellcolor{lightgray}1.318 & \cellcolor{lightgray}1.261 & \cellcolor{lightgray}1.246 & \cellcolor{lightgray}1.473  & \cellcolor{lightgray}1.151 & \cellcolor{lightgray}1.204 & \cellcolor{lightgray}1.171 & \cellcolor{lightgray}1.148 & \cellcolor{lightgray}1.278 & \cellcolor{lightgray}1.068\\
     \cline{2-15}
 & \emph{Improv.} & $f$ & 1.267 &  1.099 & 1.279 & 1.223 & 1.208 & 1.429 & 1.116 & 1.168 & 1.136 & 1.114 & 1.240 & 1.035 \\
 & 3\% &$MSE_k$  & 1.270 & 1.133 & 1.314 & 1.248 & 1.246& 1.453  & 1.029 & 1.077& 1.047 & 1.022& 1.186 & 0.961\\
    \cline{2-15}
 &  \emph{Improv.}  & $f$ & 1.241 &  1.076 & 1.252 & 1.198 & 1.183 & 1.400 & 1.093 & 1.144 & 1.113 & 1.091 & 1.214 & 1.014  \\
  & 5\% & $MSE_k$  & 1.257 & 1.133 & 1.308 & 1.245 & 1.245 &   1.437 & 1.025 & 1.060 & 1.032& 1.004& 1.165 & 0.945 \\
     \cline{2-15}
 & \emph{Improv.} & $f$ & 1.215 & 1.053 & 1.226  & 1.173 & 1.158 & 1.370 & 1.070 & 1.119 & 1.089 & 1.068 & 1.189 & 0.993 \\
 & 7\% &$MSE_k$  & 1.246 & 1.133 & 1.312 & 1.241 & 1.246 &  1.425 & 1.018& 1.042 & 1.023 & 0.987& 1.144 & 0.929 \\
    \cline{2-15}
 &  \emph{Improv.} & $f$  & 1.176 & 1.019 & 1.186 & 1.135 & 1.121 & 1.326& 1.036 &  1.083 & 1.054 & 1.033 & 1.150 & 0.961  \\
 & 10\% &$MSE_k$   & 1.231 & 1.132 & 1.313 & 1.230 & 1.246 &   1.401 & 1.005& 1.016 & 1.006 & 0.961 & 1.114& 0.905 \\
    \cline{2-15}
&  \emph{Improv.} & $f$ & 1.110 & 0.963 & 1.120 & 1.072 & 1.059 & 1.252 & 0.978 & 1.023 & 0.995 & 0.976 & 1.086 & 0.907 \\
& 15\% &$MSE_k$   & 1.207 & 1.114 & 1.271 & 1.193 & 1.241 &  1.376 & 0.971& 0.977& 1.039& 0.923& 1.060 & 0.866 \\
   \cline{2-15}
&  \emph{Improv.} & $f$ & 1.045 & 0.906 & 1.054 & 1.009 & 0.996 & 1.179 & 0.921 & 0.963 & 0.937 & 0.919 & 1.022 & 0.854 \\
& 20\% &$MSE_k$  & 1.184 & 1.081 & 1.266 & 1.150 & 1.230 & 1.354 & 0.952 & 0.972 & 0.995& 0.908 & 1.057 & 0.857 \\
 \hline
\end{tabular}
\caption{Median errors over test sets for synthetic datasets}
\label{tabla_Simulados}
\end{table*}

In particular, the values of $f$ have been set as improvement percentages over the Lasso values, where the improvement levels are $3\%$, $5\%$, $7\%$, $10\%$, $15\%$ and $20\%$ ($\gamma$ equal to $0.03$, $0.05$, $0.07$, $0.15$ and $0.20$, respectively). The results are obtained for different combinations $(p, n_k)$, where, as commented before, $p$ is chosen from the set $\{20, 100, 500\}$ and $n_k$ from $\{150,500\}$. For example, if $n_k=150$ and $p=20$, the median of the mean squared error for the Lasso in \emph{Group 1} was equal to 1.084. If the goal is to achieve an improvement of $3\%$, $f$ must be chosen equal to $1.052$. The median of the mean squared errors for the CSCLasso is equal to $0.933$ in this case (results obtained on the test sample). It is important to remark that, for some levels of improvement, the CSCLasso problem is unfeasible due to the fact that $\gamma_{max}$ for such datasets is smaller than the required $\gamma$, and, therefore, those cases are represented as empty spaces in the table. It must be noted that $\gamma_{max}$ will also depend on each fold in the cross-validation because it is associated with the partition of the data, since the $MSE_l$ in (\ref{gamma_max}) depends on such partition ($l=1,\ldots,L$). It is also worth mentioning that the constraints will always be  satisfied on the training set but not necessarily on the test set, see, for example, the case $k=3$, $n_k=150$, $p=100$ with improvement level equal to $15\%$. This phenomenon is particularly common as $p$ increases and $n_k$ decreases (see for example the values corresponding to $p=500$ and $n_k=150$ in Table \ref{tabla_Simulados}).

\begin{figure}[h!]
\centering
\subfigure{\includegraphics[width=113mm]{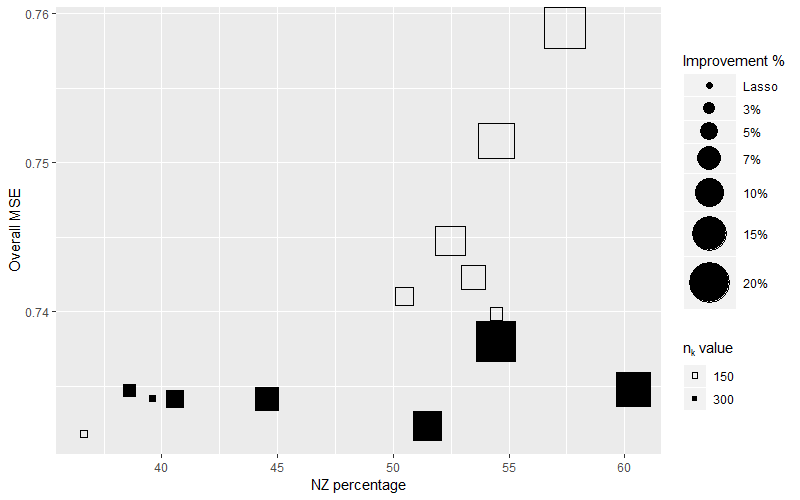}}
\caption{Median overall MSE over the test sets and NZ percentage under the choice $p=100$}
\label{fig:NZp100}
\end{figure}

We next investigate how the improvement in the prediction errors of the groups of interest affects the prediction errors in the rest of the groups, the overall prediction error and the sparsity level. Figure \ref{fig:NZp100} represents the percentage of non-zero (NZ) coefficients and the overall prediction error for different sample sizes, different levels of improvement and for $p=100$.  Lasso results are also included. For instance, when $n_k=300$ (black squares in Figure \ref{fig:NZp100}), the NZ percentage for Lasso is 39.60 with an associated overall MSE of 0.734; whereas running the CSCLasso demanding a 3\% of improvement over the first six groups, we achieve a NZ percentage of 38.61, and an overall MSE of 0.735. In general terms, it can be seen that the sparsity of the solution decreases with the improvement level: smaller squares, which represent smaller imposed improvement percentages, are on the left of bigger squares (which are associated with demanding percentage of improvement). Then, if the user is very demanding in predicting a specific group, this implies, in the majority of the cases, a less sparse solution. Notwhitstanding, when no level of improvement is imposed (Lasso problem), the solution can be less sparse than in the CSCLasso, as in the case of $n_k=300$. This also occurs when $p=500$ and $n_k=150$ (see the bottom graphic of Figure \ref{fig:NZp20and500} in Appendix: further results). Furthermore, the overall prediction error slightly worsens with the improvement level, due to the worsening of the predictions in the uncontrolled groups.

\begin{figure}[h!]
\centering \small
\subfigure{\includegraphics[width=112mm]{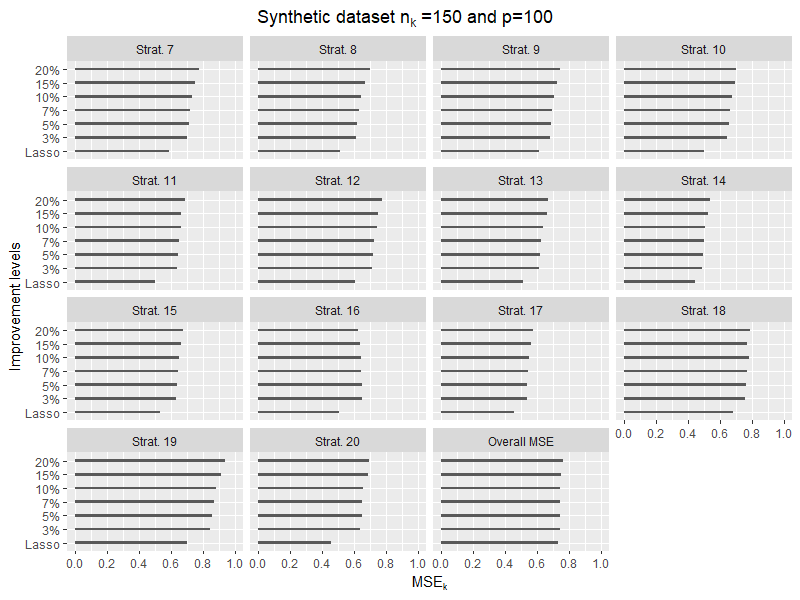}}
\subfigure{\includegraphics[width=112mm]{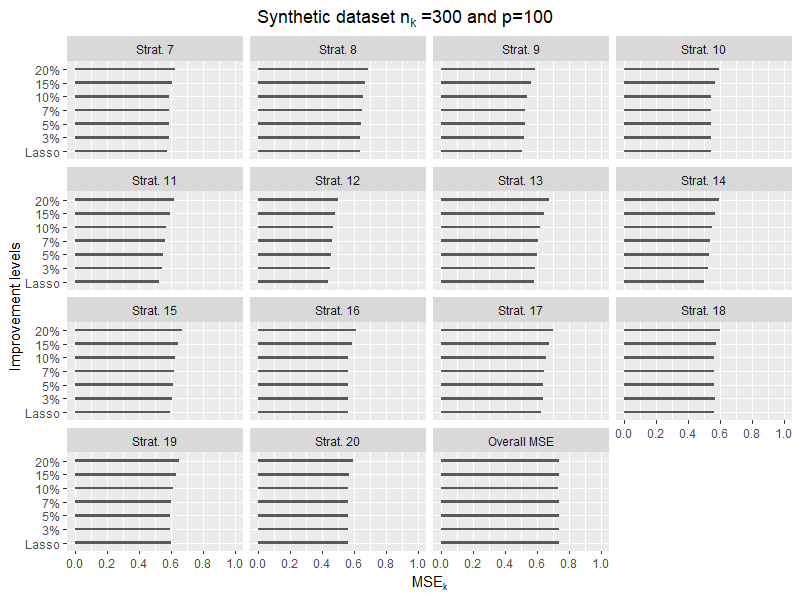}}
\caption{Median $MSE_k$ over the test sets for $k=7,\ldots,20$ under $p=100$ features and, $n_k=150$ (top) and $n_k=300$ (bottom). Each subgraph represents one group and the Y-axis shows the different percentages of improvement}
\label{fig:p100}
\end{figure}

Figure \ref{fig:p100} represents the prediction errors over the groups that are not controlled as well as the overall mean squared error, for different improvement levels, $n_k=150$ (top figure) and $n_k=300$ (bottom) when $p=100$. In the figure, Lasso values are also shown. From the figure, it can be concluded that the Lasso performs better in the uncontrolled groups, since the prediction errors worsen under our proposal. However, the overall mean squared error remains almost constant (since the improved errors compensate the more deteriorated ones). Similar conclusions can be drawn under the choices $p=20$ and $p=500$ (see Figures \ref{fig:p20} and \ref{fig:p500} in Appendix: further results, respectively).

Next, we test how the solution of the CSCLasso behaves with respect to other global performance measures as the $l_2$ distance, false positive and negative rates, which are defined not in terms of prediction errors, but on the correct fitting of the generator process (see \cite{doi:10.1080/01621459.2015.1034319}). In particular, the $l_2$ distance is defined as $\|\boldsymbol{\hat{\beta}}-\boldsymbol{\beta}\|_2$, where $\boldsymbol{\beta}$ is the vector of coefficients that generated the datasets (described at the beginning of this section), and $\boldsymbol{\hat{\beta}}(\lambda)$ are the estimators. In addition, the false positive rate (FPR) and false negative rate (FNR) are calculated as follows:
\begin{equation*}
\text{FPR} = \frac{\mid j : \beta_j=0 \,\, \& \,\, \hat{\beta}_j(\lambda)=0\mid}{\mid j : \beta_j=0\mid},
\end{equation*}

\begin{equation*}
\text{FNR} = \frac{\mid j : \beta_j\neq0 \,\, \& \,\, \hat{\beta}_j(\lambda)=0\mid}{\mid j : \beta_j\neq0\mid},
\end{equation*}
where $j=1,\ldots,p$. The median of these three measures as well as the median of the overall MSE (already shown in Figures \ref{fig:NZp100}-\ref{fig:p100} and Figures \ref{fig:p20}, \ref{fig:p500} and \ref{fig:NZp20and500} in Appendix: further results), are presented in Table \ref{tabla_Simulados_othermeasures}. For the choices where $p=20$, the FPR values are not given since all the predictors have associated non-zero coefficients when the datasets were created. From this table it can be deduced that similar or even better results, comparing with those of the Lasso, are obtained in the majority of the cases across the four different measures.

\begin{table*}[h!]
\centering
\small
\hspace*{-1.1cm}\begin{tabular}{clcccc|cccc}
\cline{3-10}
  & &\multicolumn{4}{c}{$n_k=150$} & \multicolumn{4}{c}{$n_k=300$}  \\
  \hline
$p$ & & Overall MSE & $l_2$ distance & FPR & FNR & Overall MSE & $l_2$ distance & FPR & FNR \\
\hline
  \multirow{7}{*}{20} & \cellcolor{lightgray}\emph{Lasso}  & \cellcolor{lightgray}0.667 & \cellcolor{lightgray}4.149 & \cellcolor{lightgray}- & \cellcolor{lightgray}0.050 &\cellcolor{lightgray}0.666  & \cellcolor{lightgray}4.146 & \cellcolor{lightgray}- & \cellcolor{lightgray}0.000 \\
 \cline{2-10}
&  \emph{Improv.} 3\% & 0.691  & 4.108  & - & 0.000 &0.673  & 4.099 & - &0.000\\
  \cline{2-10}
& \emph{Improv.} 5\%& 0.695  & 4.104  &  - & 0.000 & 0.683 & 4.089 & - & 0.000 \\
  \cline{2-10}
 & \emph{Improv.} 7\% & -& - & - & - &0.693 & 4.077 & - & 0.000\\
   \cline{2-10}
 &\emph{Improv.} 10\%  &-& - &-&-& - & -& -& -\\
    \cline{2-10}
 & \emph{Improv.} 15\% &-& - &-&-& - & -& -& -\\
  \cline{2-10}
 & \emph{Improv.} 20\%  &-& - &-&-& - & -& -& -\\
 \hline
   \multirow{7}{*}{100}  & \cellcolor{lightgray}\emph{Lasso}  & \cellcolor{lightgray}0.732 & \cellcolor{lightgray}2.931 & \cellcolor{lightgray}0.275 & \cellcolor{lightgray}0.250 &\cellcolor{lightgray} 0.734 & \cellcolor{lightgray}2.729 & \cellcolor{lightgray}0.250 & \cellcolor{lightgray}0.150  \\
 \cline{2-10}
&  \emph{Improv.} 3\% &  0.740 & 2.857 & 0.388 & 0.100 & 0.735 & 2.721 & 0.163 & 0.150\\
  \cline{2-10}
& \emph{Improv.} 5\%&  0.741 &  2.855 & 0.375  & 0.100 & 0.734  & 2.721 &  0.163 & 0.100\\
  \cline{2-10}
 & \emph{Improv.} 7\% & 0.742& 2.853 & 0.400 & 0.100 & 0.734 & 2.718& 0.163 & 0.100 \\
   \cline{2-10}
 &\emph{Improv.} 10\%  & 0.745& 2.849 & 0.400 & 0.100& 0.732 & 2.704& 0.275 & 0.100 \\
    \cline{2-10}
 & \emph{Improv.} 15\% &0.751 & 2.844& 0.425 & 0.100&0.735 &2.678 & 0.288 & 0.050\\
  \cline{2-10}
 & \emph{Improv.} 20\%  & 0.759 & 2.835 & 0.463 & 0.100& 0.738 &2.655 &0.288 & 0.100\\
 \hline
     \multirow{7}{*}{500} &  \cellcolor{lightgray}\emph{Lasso} & \cellcolor{lightgray}0.772 & \cellcolor{lightgray}2.778 & \cellcolor{lightgray}0.135 & \cellcolor{lightgray}0.300 &\cellcolor{lightgray}0.744  & \cellcolor{lightgray}2.750 & \cellcolor{lightgray}0.065 & \cellcolor{lightgray}0.300  \\
 \cline{2-10}
&  \emph{Improv.} 3\% &  0.772 & 2.779  & 0.040 & 0.350 &0.744  & 2.658 & 0.063 & 0.250\\
  \cline{2-10}
& \emph{Improv.} 5\%&  0.772 &  2.775 &  0.050 & 0.300 & 0.745 & 2.653 &  0.075 & 0.250 \\
  \cline{2-10}
 & \emph{Improv.} 7\% & 0.772 & 2.769 & 0.042 & 0.350 &0.746 &2.647& 0.077 & 0.250\\
   \cline{2-10}
 &\emph{Improv.} 10\%  & 0.772 & 2.760 & 0.056 & 0.350& 0.748 &2.632 & 0.104 & 0.250 \\
    \cline{2-10}
 & \emph{Improv.} 15\% & 0.771 & 2.754 & 0.063 & 0.300 & 0.752 &2.640 & 0.129 & 0.250\\
  \cline{2-10}
 & \emph{Improv.} 20\%  & 0.774 & 2.735 & 0.069 & 0.350& 0.760& 2.607 &0.148 & 0.200\\
 \hline
\end{tabular}
\caption{Median performance measures over test sets for synthetic datasets}
\label{tabla_Simulados_othermeasures}
\end{table*}

\begin{figure}[h!]
\centering
\includegraphics[width=110mm]{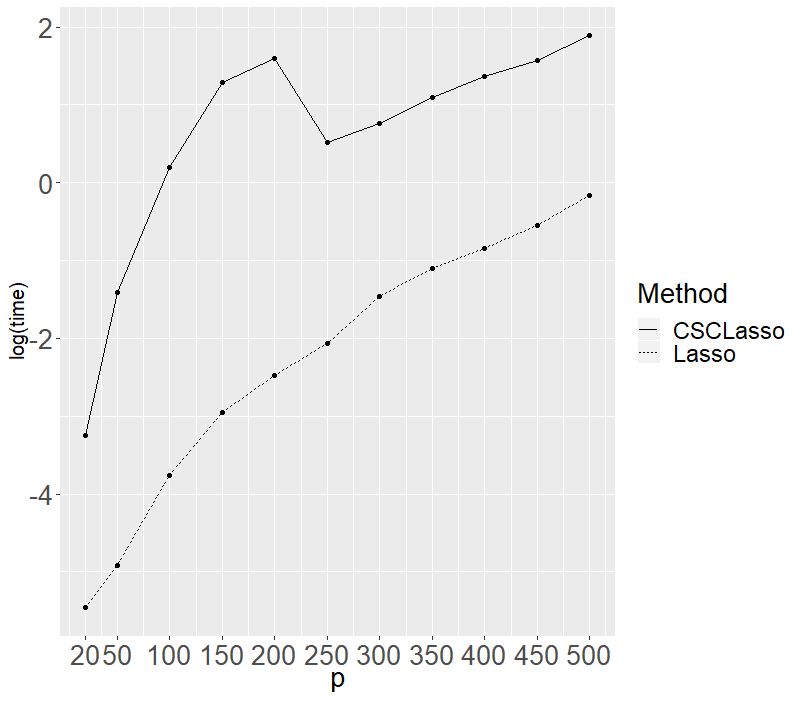}
\caption{A two-dimensional graph of the logarithm of the user times in seconds for $n_k=300$ as $p$ increases}
\label{fig:times_p500}
\end{figure}

A final remark concerning the computational cost of the CSCLasso when comparing with Lasso is as follows. The median user time required to solve the problem with the largest dataset considered in this study ($n_k=300$ and $p=500$) is $0.85$ seconds when the Lasso is run on Intel(R) Core(TM) i7-7500U CPU at 2.70GHz 2.90GHz with 8.0 GB of RAM; whereas the CSCLasso requires $6.60$ seconds. Nevertheless, to better understand how the computation time behaves depending on $p$ value, a grid in this parameter has been inspected, while $n_k$ is set to 300. Figure \ref{fig:times_p500}
depicts the logarithm of the user times in seconds obtained under Lasso and CSCLasso models when $n_k=300$ and $p$ changes. Then, under a reasonable computational cost, the desired results are achieved. Further analyses regarding the computational times are shown in the Appendix (Figure \ref{fig:times}).


\subsection{Leukemia dataset: a gene expression dataset}\label{realdataset}

The real stratified dataset described in \cite{Kouno2013} is explored here. The data contain information related to myeloid monocytic leukemia cells undergoing differentiation to macrophages. In particular, the dataset is formed by expression levels of $45$ transcription factors (response and predictor variables) measured at $8$ distinct times (groups) of the differentiation process. As in \cite{doi:10.1093/biomet/asw065}, the aim is to predict the EGR2 transcription factor in terms of the other $p=44$ factors. The sample size per group is equal to $120$. Similarly as in Section \ref{simulateddatasets}, the Lasso was run and the overall prediction errors, individual prediction errors per group and percentage of non-zero coefficients are recorded. The records in \emph{Group 1} yield the best MSE using Lasso model. Therefore, we may be interested in obtaining an even better fitting for such data. The CSCLasso problem is solved with threshold values smaller than the Lasso error, which turns out to be 0.370. Table \ref{tabla_gene} shows the obtained median results for an assortment of improvement levels, namely, $5\%$, $7\%$, $10\%$ and $15\%$ or, equivalently, $\gamma$ is equal to $0.05$, $0.05$, $0.07$ and $0.15$, respectively. From that table, it can be seen how the prediction error of interest (corresponding to \emph{Group 1}) decreases with the improvement level, as expected. Similarly as in Section \ref{simulateddatasets}, the overall mean squared error does not exhibit significant changes, while the prediction errors in the uncontrolled groups do not exhibit the same behaviour. Some of them slightly improve (\emph{Group 6}), others slightly worsen (as the \emph{Group 5} and \emph{Group 8}) and others remain constant (as \emph{Group 2}). Finally, in regards to the sparsity of the solution, for this dataset, less sparse solutions are obtained by CSCLasso in comparison with the Lasso ones.

\begin{table*}[h!]
\centering \small
\hspace*{-1cm}\begin{tabular}{lccccccccccc}
\cline{2-12}
& $f_1$ & \emph{Overall MSE} & $MSE_1$& $MSE_2$ & $MSE_3$ & $MSE_4$ & $MSE_5$ & $MSE_6$ & $MSE_7$ & $MSE_8$ & NZ\\
\hline

\cellcolor{lightgray}\emph{Lasso} & \cellcolor{lightgray}- & \cellcolor{lightgray} 0.620 & \cellcolor{lightgray} $\boldsymbol{0.370}$ & \cellcolor{lightgray}0.417& \cellcolor{lightgray}0.902& \cellcolor{lightgray}0.496 & \cellcolor{lightgray}0.480 & \cellcolor{lightgray}0.535 & \cellcolor{lightgray} 0.496 & \cellcolor{lightgray}0.685 &\cellcolor{lightgray} 53.33\\ \hline
  \emph{Improv.} 5\% & $\boldsymbol{0.352}$&  0.623 & $\boldsymbol{0.357}$  & 0.417 &	0.918&	0.500 &	0.555 &	0.512 &	0.497 &	0.719 &	73.33
\\ 
 \emph{Improv.} 7\% & $\boldsymbol{0.344}$& 0.626  & $\boldsymbol{0.348}$  & 0.418 &	0.915 &	0.504 &	0.565 &	0.514 &	0.497 &	0.722 &	66.67
\\
 \emph{Improv.} 10\% & $\boldsymbol{0.333}$&  0.630 & $\boldsymbol{0.335}$  & 0.418 &	0.911 &	0.510&	0.574 &	0.513 &	0.498 &	0.728 & 66.67
\\
 \emph{Improv.} 15\% & $\boldsymbol{0.315}$& 0.636  & $\boldsymbol{0.331}$  & 0.415 &	0.903 &	0.523 &	0.591 &	0.512 &	0.502 &	0.737 &	73.33
\\
 \hline
\end{tabular}
\caption{Median errors over test set for gene expression dataset. Constraints imposed over \emph{Group 1}}
\label{tabla_gene}
\end{table*}

\subsection{Communities and Crime dataset}
In this section, a real dataset from the UCI Machine Learning Repository (\cite{Lichman:2013}) will be analyzed. In particular, the so-called \emph{Communities and Crime Unnormalized Data Set} shall be considered. The dataset is about communities within the United States and has already been inspected in the literature (see \cite{REDMOND2002660}).
This dataset combines crime information from the FBI databases \cite{crimedataset3} as well as socio-economic and law enforcement data from \cite{crimedataset1} and \cite{crimedataset2}, respectively. The dataset is formed by $p=124$ predictors, $23$ of which present missing values, and $n=2215$ instances, where the response variable measures the number of murders per 100K population. The predictor variables with missing values are not consider for the next experiments. As such, we finally consider $p=101$ predictors. Additionally, for each instance (community), the region from which it comes is known. Thus, if we were interested in obtaining a good prediction in a certain region, say Midwest, we could control these communities by including a performance constraint. Table \ref{tabla_Communities_Crime} shows the median errors over the test set for \emph{Group 1}, formed by the communities of Midwest, and over the rest of communities (\emph{Group 2}). In terms of overall MSE and MSE over the two groups, similar conclusions as in Section \ref{realdataset} are drawn. Whereas different improvement levels are imposed, the MSE of interest ($MSE_1$) is getting smaller but the overall prediction error is almost not affected by the constraint. Lastly, regarding the sparsity of the solution, an analogous behaviour as that observed in the case of simulated data is obtained: the solution becomes less sparse with the improvement level.

\begin{table*}[h!]
\centering \small
\hspace*{-1cm}\begin{tabular}{lccccc}
\cline{2-6}
& $f_1$ & \emph{Overall MSE} & $MSE_1$& $MSE_2$  & NZ\\
\hline

\cellcolor{lightgray}\emph{Lasso} & \cellcolor{lightgray}- & \cellcolor{lightgray}0.488 & \cellcolor{lightgray}  $\boldsymbol{0.433}$ & \cellcolor{lightgray} 0.453& \cellcolor{lightgray} 21.57 \\ \hline
 \emph{Improv.} 5\% & $\boldsymbol{0.411}$& 0.488  & $\boldsymbol{0.422}$  & 0.453 &	25.49 \\
 \emph{Improv.} 7\% & $\boldsymbol{0.403}$&  0.487 & $\boldsymbol{0.420}$  & 0.453 &	28.43 \\
 \emph{Improv.} 10\% & $\boldsymbol{0.390}$&  0.488 & $\boldsymbol{0.416}$  & 0.453 &	26.47 \\
 \emph{Improv.} 15\% & $\boldsymbol{0.368}$&  0.486 & $\boldsymbol{0.403}$  &  0.459 &	34.31 \\
 \hline
\end{tabular}
\caption{Median errors over test set for communities and crime dataset. Constraints imposed over \emph{Group 1}}
\label{tabla_Communities_Crime}
\end{table*}

As previously commented, the groups of interest may overlap. As an illustration, assume that the interest is in controlling the prediction error in communities of Midwest or communities with a population density larger than or equal to the 75th percentile. Let \emph{Group 1} denote the communities from Midwest, while \emph{Group 2} represents the communities where the density of population is higher than the 75th percentile. For instance, if we aim to improve in a $7\%$ the errors obtained by the Lasso model (equal to $0.513$ and $0.442$), then the CSCLasso results become $0.475$ and $0.441$, respectively.

\section{Conclusions and extensions}\label{Conclusions}
In this paper a new version of the Lasso regression model that strives to control the performance rates associated with individuals of interest is proposed. The method has
a significant application in the context of heterogeneous data, where it is common that
certain sources are more reliable than others, or simply the prediction on some groups of data are of higher interest, and thus a better fit is sought for some data. In order to control the individuals of interest, performance constraints are included in the regression model. This approach leads to a novel method (CSCLasso) which is not reported in the literature previously, up to our knowledge. Theoretical results concerning this novel methodology have been discussed and, in addition, the CSCLasso has been tested on six synthetic datasets with different properties, on a well-referenced real stratified biomedical dataset and on a real social sciences dataset. The numerical section shows that, with a low computational cost, the accuracy prediction errors for the groups of interest are controlled. This is done at the expense of reducing sparsity (if the regularization parameter is kept fixed) or the overall accuracy.

A number of extensions to this work are possible. Regarding the selection of the threshold values $f_1,\ldots,f_L$, as commented in Section \ref{role}, a more flexible choice would be to consider a different fine-tuning parameter $\tau$ (or $\gamma$) for each group of interest $l$, say $\tau_l$ ($\gamma_l$). However, this generalization implies tuning (many) more parameters, making the model less usable by users. In addition, for the sake of dealing with strongly correlated predictors, it may be of interest to change the objective function by that of the elastic net (\cite{elasticnet}). Another non-straightforward extension could be to address classification problems (via the logistic regression) instead of regression problems. In this case, we would not adress a quadratic problem. Work in these issues is underway.

\section*{Acknowledgements}
Research partially supported by research grants and projects MTM2015-65915-R (Ministerio de Econom\'ia y Competitividad, Spain), FQM-329 and P18-FR-2369 (Junta de Andaluc\'ia, Spain), Fundaci\'on BBVA and EC H2020 MSCA RISE NeEDS Project (Grant agreement ID: 822214). In addition, we would like to thank the associated editor and two anonymous reviewers for carefully reading this work, and for their insightful comments, which have helped to improve the quality of this paper.

\section*{Appendix: proofs}
\subsection*{Proof of Proposition \ref{Prop_1}}
Given $\lambda\geq 0$, consider problem (\ref{CSCLassoOneConst}). If $\boldsymbol{\beta} = \boldsymbol{\beta}^+ - \boldsymbol{\beta}^-$ with $\boldsymbol{\beta}^+ \geq 0$ and $\boldsymbol{\beta}^- \geq 0$ and $\boldsymbol{\lambda}=(0,\lambda,\ldots,\lambda)'$, a vector whose length is $p+1$, then the differentiable version of that problem turns out to be

\begin{align*}
\begin{split}
\underset{\boldsymbol{\beta^+,\beta^-}}{\min} \hspace{0.5cm}& \dfrac{1}{n_0}\| \mathbf{y}_0-\mathbf{X}_0(\boldsymbol{\beta}^+ - \boldsymbol{\beta}^-)\|^{2} + \boldsymbol{\lambda}'\boldsymbol{\beta}^+ + \boldsymbol{\lambda}'\boldsymbol{\beta}^-\\
\mbox{s.t.} \hspace{0.5cm} & \dfrac{1}{n_1}\| \mathbf{y}_1-\mathbf{X}_1(\boldsymbol{\beta}^+ - \boldsymbol{\beta}^-)\|^{2} -(1+ \tau)MSE_1(\boldsymbol{\hat{\beta}}^{ols}) \leq 0,\\
& \boldsymbol{\beta}^+ \geq \boldsymbol{0} \Leftrightarrow -\boldsymbol{\beta}^+ \leq \boldsymbol{0},\\
&  \boldsymbol{\beta}^- \geq \boldsymbol{0} \Leftrightarrow -\boldsymbol{\beta}^- \leq \boldsymbol{0}.
\label{}\end{split}
\end{align*}
From the Karush-Kuhn-Tucker conditions,
$$\mathcal{L}(\boldsymbol{\beta}^+, \boldsymbol{\beta}^-, \boldsymbol{\theta}^+, \boldsymbol{\theta}^-, \eta)= \dfrac{1}{n_0}\|\mathbf{y}_0-\mathbf{X}_0(\boldsymbol{\beta}^+-\boldsymbol{\beta}^-)\|^2 + \boldsymbol{\lambda}'\boldsymbol{\beta}^+ + \boldsymbol{\lambda}'\boldsymbol{\beta}^- - (\boldsymbol{\theta}^+)' \boldsymbol{\beta}^+ - (\boldsymbol{\theta}^-)'\boldsymbol{\beta}^- + $$
$$+ \eta\left(\dfrac{1}{n_1}\|\mathbf{y}_1 -\mathbf{X}_1(\boldsymbol{\beta}^+-\boldsymbol{\beta}^-)\|^2 - (1+\tau)MSE_1(\boldsymbol{\hat{\beta}}^{ols})\right)$$

$$\dfrac{\partial}{\partial \boldsymbol{\beta}^+}: -\dfrac{2}{n_0}\mathbf{X}_0'(\mathbf{y}_0-\mathbf{X}_0(\boldsymbol{\beta}^+ - \boldsymbol{\beta}^-)) + \boldsymbol{\lambda} - \boldsymbol{\theta}^+ - \dfrac{2}{n_1}\eta \mathbf{X}_1'(\mathbf{y}_1 - \mathbf{X}_1(\boldsymbol{\beta}^+ - \boldsymbol{\beta}^-)) = 0$$
$$\dfrac{\partial}{\partial \boldsymbol{\beta}^-}: \dfrac{2}{n_0}\mathbf{X}_0'(\mathbf{y}_0-\mathbf{X}_0(\boldsymbol{\beta}^+ - \boldsymbol{\beta}^-)) + \boldsymbol{\lambda} - \boldsymbol{\theta}^- + \dfrac{2}{n_1}\eta \mathbf{X}_1'(\mathbf{y}_1 - \mathbf{X}_1(\boldsymbol{\beta}^+ - \boldsymbol{\beta}^-)) = 0$$
$$\boldsymbol{\theta}^+, \boldsymbol{\theta}^-, \eta \geq 0$$
$$(\boldsymbol{\theta}^+)' \boldsymbol{\beta}^+ = 0$$
$$(\boldsymbol{\theta}^-)' \boldsymbol{\beta}^- = 0$$
$$\eta \left(\dfrac{1}{n_1}\|\mathbf{y}_1 - \mathbf{X}_1(\boldsymbol{\beta}^+ - \boldsymbol{\beta}^-)\|^2 -(1+\tau)MSE_1(\boldsymbol{\hat{\beta}}^{ols})\right) = 0$$
Thus,
\begin{itemize}
\item if $\boldsymbol{\beta}>0 \Rightarrow \boldsymbol{\beta}^+>0, \boldsymbol{\beta}^- =0 \Rightarrow \boldsymbol{\theta}^+=0 \Rightarrow  -\dfrac{2}{n_0}\mathbf{X}_0'(\mathbf{y}_0-\mathbf{X}_0\boldsymbol{\beta}) + \boldsymbol{\lambda} -\dfrac{2}{n_1}\eta \mathbf{X}_1'(\mathbf{y}_1 - \mathbf{X}_1\boldsymbol{\beta}) = 0$
\item if $\boldsymbol{\beta}<0 \Rightarrow \boldsymbol{\beta}^+=0, \boldsymbol{\beta}^- >0 \Rightarrow \boldsymbol{\theta}^-=0 \Rightarrow  \dfrac{2}{n_0}\mathbf{X}_0'(\mathbf{y}_0-\mathbf{X}_0\boldsymbol{\beta}) + \boldsymbol{\lambda} +\dfrac{2}{n_1}\eta \mathbf{X}_1'(\mathbf{y}_1 - \mathbf{X}_1\boldsymbol{\beta}) = 0$
\end{itemize}
Therefore,
\begin{equation}
\label{eq:kkt}
\dfrac{2}{n_0}\mathbf{X}_0'(\mathbf{y}_0-\mathbf{X}_0\boldsymbol{\beta}) +\dfrac{2}{n_1}\eta(\lambda) \mathbf{X}_1'(\mathbf{y}_1 - \mathbf{X}_1\boldsymbol{\beta}) = \mathbf{b}(\lambda),
\end{equation}
where $\eta(\lambda)$ is the Lagrange multiplier associated with the first constraint and $\mathbf{b}(\lambda)$ is a $(p+1)$-dimensional vector whose $s$-th component, $s=0,1,\ldots,p$, takes the following value
$$b_s(\lambda)=\left\{ \begin{array}{lcc}
              \,\,\,\,\, \lambda,  \,\, \text{if}  \,\,  \beta_s>0, \\
             -\lambda, \,\,  \text{if} \,\,  \beta_s<0, \\
             \,\,\,\,\, 0,  \,\, \,\, \text{else.}
             \end{array}\right.$$
             
Then, since $\mathbf{X}_0$ and $\mathbf{X}_1$ are maximum rank matrices, one obtains from (\ref{eq:kkt}) the following implicit expression for the solution
$\boldsymbol{\hat{\beta}}^{CSCLasso}(\lambda)$ of Problem (\ref{CSCLassoOneConst})
$$\boldsymbol{\hat{\beta}}^{CSCLasso}(\lambda)= \left(\dfrac{1}{n_0}\mathbf{X}_0'\mathbf{X}_0 + \dfrac{1}{n_1}\eta(\lambda)\mathbf{X}_1'\mathbf{X}_1\right)^{-1}\left(\dfrac{1}{n_0}\mathbf{X}_0'\mathbf{y}_0 + \dfrac{1}{n_1}\eta(\lambda)\mathbf{X}_1'\mathbf{y}_1\right) - \dfrac{1}{2}\left(\dfrac{1}{n_0}\mathbf{X}_0'\mathbf{X}_0 + \dfrac{1}{n_1}\eta(\lambda)\mathbf{X}_1'\mathbf{X}_1\right)^{-1} \mathbf{b}(\lambda). $$

\subsection*{Proof of Theorem \ref{TheoExisandUniq}}

Consider the function $h:\boldsymbol{\beta} \mapsto \dfrac{1}{n}\| \mathbf{y} - \mathbf{X}\boldsymbol{\beta}\|^2=\dfrac{1}{n}(\mathbf{y}-\mathbf{X}\boldsymbol{\beta})'(\mathbf{y}-\mathbf{X}\boldsymbol{\beta})$, where $\mathbf{X}$ is a maximum rank matrix by hypothesis. The matrix $\mathbf{X}$ is of maximum rank and therefore the Hessian matrix $H_h(\boldsymbol{\beta})=\dfrac{2}{n}\mathbf{X}'\mathbf{X}$ is positive definite, from where we conclude that $h(\boldsymbol{\beta})$ is strictly convex, and hence,
$h(\boldsymbol{\beta}) +  \lambda \| \mathcal{A}\boldsymbol{\beta}\|_1$
is also a strictly convex function.

We next prove that $h(\boldsymbol{\beta})$ is a coercive function. Since $\mathbf{X}'\mathbf{X}$ is positive definite, its eigenvalues are all positive. In particular, the smallest eigenvalue, say $\gamma_r$, will be nonzero. Moreover, using the spectral decomposition of a symmetric matrix,

$$\dfrac{1}{n}\| \mathbf{y} - \mathbf{X}\boldsymbol{\beta}\|^2=\dfrac{1}{n}(\mathbf{y}-\mathbf{X}\boldsymbol{\beta})'(\mathbf{y}-\mathbf{X}\boldsymbol{\beta})=\dfrac{1}{n}\boldsymbol{\beta}'\mathbf{X}'\mathbf{X}\boldsymbol{\beta} - \dfrac{2}{n}\mathbf{y}'\mathbf{X}\boldsymbol{\beta} + \dfrac{1}{n}\mathbf{y}'\mathbf{y}=$$
$$ =\dfrac{1}{n}\boldsymbol{\beta}'Q'DQ\boldsymbol{\beta}-\dfrac{2}{n}\mathbf{y}'\mathbf{X}\boldsymbol{\beta}+ \dfrac{1}{n}\mathbf{y}'\mathbf{y} \geq$$
$$\geq \dfrac{1}{n}\boldsymbol{\beta}'Q'DQ\boldsymbol{\beta}-\left| \dfrac{2}{n}\mathbf{y}'\mathbf{X}\boldsymbol{\beta}\right|+ \dfrac{1}{n}\mathbf{y}'\mathbf{y} \geq \dfrac{\gamma_r}{n}\| Q\boldsymbol{\beta}\|^2 - \left\|\dfrac{2} {n}\mathbf{y}'\mathbf{X}\right\| \| \boldsymbol{\beta}\| +\dfrac{1}{n} \mathbf{y}'\mathbf{y} = $$
$$=\dfrac{\gamma_r}{n}\| \boldsymbol{\beta}\|^2 - \left\|\dfrac{2}{n}\mathbf{y}'\mathbf{X}\right\| \| \boldsymbol{\beta}\| + \dfrac{1}{n} \mathbf{y}'\mathbf{y},$$
where, in the second-to-last step, the Cauchy-Schwarz inequality has been used. As $\| \boldsymbol{\beta}\| \rightarrow +\infty$, then $h(\boldsymbol{\beta})\rightarrow +\infty$ too, and thus $h(\boldsymbol{\beta})$ is a coercive function.

Now we show that (\ref{CSCLasso_GeneralB}) has optimal solution. Let $\boldsymbol{\beta}^*\in \boldsymbol{B}$. As $h(\boldsymbol{\beta})$ is coercive, then there exists $R>0$ such that $$\dfrac{1}{n}\| \mathbf{y} - \mathbf{X}\boldsymbol{\beta}\|^2 \,\,>  \dfrac{1}{n}\| \mathbf{y} - \mathbf{X}\boldsymbol{\beta}^*\|^2 + \lambda \| \mathcal{A}\boldsymbol{\beta}^*\|_1,$$
for all $\boldsymbol{\beta}$ such that $\| \boldsymbol{\beta}\| > R$. For that reason, the problem can be reduced to the feasible compact region
$\boldsymbol{B}\cap\{\boldsymbol{\beta}: \,\, \|\boldsymbol{\beta}\|  \leq  R\}$, which implies that the optimal solution is reached. Finally, the uniqueness of the solution follows from the fact that $h(\boldsymbol{\beta}) + \lambda\|\mathcal{A}\boldsymbol{\beta}\|_1$ is strictly convex.

\subsection*{Proof of Proposition \ref{Prop_2}}
Let us consider the optimization problem (\ref{lassoA}) and let $\boldsymbol{\hat{\beta}}^{Lasso}(\lambda)$ denotes its optimal solution. The necessary and sufficient optimality condition is:
\begin{equation}\label{NandSOptCond}
\nabla \dfrac{1}{n}\| \mathbf{y}-\mathbf{X} \boldsymbol{\hat{\beta}}^{Lasso}(\lambda)\|^{2} + \lambda \partial \|\mathcal{A}\boldsymbol{\hat{\beta}}^{Lasso}(\lambda)\|_1 \ni \boldsymbol{0}.
\end{equation}
From the properties of subdifferential (see \emph{Theorem 23.9} of \cite{rockafellar}) it follows that
\begin{equation*}
\partial \|\mathcal{A}\boldsymbol{\hat{\beta}}^{Lasso}(\lambda)\|_1 = \mathcal{A}' \partial\|.\|_{1_{\mid \mathcal{A}\boldsymbol{\hat{\beta}}^{Lasso}(\lambda)}},
\end{equation*}
which implies that (\ref{NandSOptCond}) becomes
\begin{equation}\label{NandSOptCond_}
-\dfrac{2}{n}\mathbf{X}'(\mathbf{y}-\mathbf{X}\boldsymbol{\hat{\beta}}^{Lasso}(\lambda)) + \lambda \mathcal{A}' \partial\|.\|_{1_{\mid \mathcal{A}\boldsymbol{\hat{\beta}}^{Lasso}(\lambda)}} \ni \mathbf{0}.
\end{equation}
Consequently, the necessary and sufficient condition (\ref{NandSOptCond_}) in $\boldsymbol{\hat{\beta}}^{Lasso}(\lambda)=\mathbf{0}$ is
\begin{equation*}
-\dfrac{2}{n}\mathbf{X}'\mathbf{y} + \lambda \{\mathcal{A}'\mathbf{t} : \|\mathbf{t}\|_{\infty} \leq 1\} \ni \mathbf{0},
\end{equation*}
since $\partial\|\mathbf{0}\|_1$ is the unit ball of the $\|.\|_{\infty}$. Equivalently,
\begin{equation*}
\dfrac{2}{n}\mathbf{X}'\mathbf{y} \in \{\mathcal{A}'\lambda \mathbf{t} : \|\mathbf{t}\|_{\infty} \leq 1\}.
\end{equation*}
Therefore, the solution of the problem

\begin{align}
\begin{split}
\underset{\lambda, \mathbf{t}}{\min} \hspace{0.5cm} & \lambda\\
\mbox{s.t.} \hspace{0.5cm} & \dfrac{2}{n}\mathbf{X}'\mathbf{y} = \mathcal{A}'\lambda \mathbf{t},\\
& \|\mathbf{t}\|_{\infty} \leq 1,\\
& \lambda \geq 0,
\label{Prob_minlambda}\end{split}
\end{align}
will provide the minimum $\lambda$ from which $\boldsymbol{\hat{\beta}}^{Lasso}(\lambda)=\mathbf{0}$ is the optimal solution.
If $\mathbf{q}=\lambda \mathbf{t}$, then Problem (\ref{Prob_minlambda}) becomes

\begin{align*}
\begin{split}
\underset{\lambda, \mathbf{q}}{\min} \hspace{0.5cm}& \lambda\\
\mbox{s.t.} \hspace{0.5cm} & \dfrac{2}{n}\mathbf{X}'\mathbf{y} = \mathcal{A}'\mathbf{q},\\
& \|\mathbf{q}\|_{\infty} \leq \lambda.\\
\label{}\end{split}
\end{align*}
The constraint $\|\mathbf{q}\|_{\infty}$ is equivalent to $|q_s| \le \lambda$, $s=0,1,\ldots,p$ and the result follows.

\subsection*{Proof of Proposition \ref{uniqueness_teo}}
The proof follows very closely that of Theorem \ref{TheoExisandUniq}. First of all, it shall be proven that  $h: \boldsymbol{\beta} \mapsto E[(Y-X'\boldsymbol{\beta})^{2}]$ is coercive. It is strictly convex on $\boldsymbol{\beta}$ since its Hessian matrix, $2E[XX']$, is positive definite due to $X$ is an absolutely continuous $p$-dimensional random variable:
$$u'E[XX']u=E[u'XX'u]=E[(X'u)^{2}]>0,$$
since $P(X'u=0)=0$. Moreover, $\lambda \| \mathcal{A}\boldsymbol{\beta}\|_1$ is a convex function on $\boldsymbol{\beta}$ and, therefore, $E[( Y-X' \boldsymbol{\beta})^{2}+ \lambda \| \mathcal{A}\boldsymbol{\beta}\|_1]$ is also a strictly convex function on $\boldsymbol{\beta}$.

On the one hand, the eigenvalues of the Hessian matrix are all positive and, in particular, the smallest eigenvalue, say $\gamma_r$, will be non-zero. On the other hand, using the spectral decomposition of a symmetric matrix,

$$E[(Y-X'\boldsymbol{\beta})^{2}]=\boldsymbol{\beta}'E[XX']\boldsymbol{\beta} - 2E[YX]\boldsymbol{\beta} +  E[Y^2]=\boldsymbol{\beta}'Q'DQ\boldsymbol{\beta} - 2E[YX]\boldsymbol{\beta} +  E[Y^2]\geq $$
$$\geq \boldsymbol{\beta}'Q'DQ\boldsymbol{\beta} - \mid 2E[YX]\boldsymbol{\beta} \mid +  E[Y^2] \geq \gamma_r\| Q\boldsymbol{\beta} \|^{2}- \| E[YX]\| \| \boldsymbol{\beta} \| + E[Y^2]=$$
$$=\gamma_r\| \boldsymbol{\beta}\|^{2}- \| E[YX]\| \| \boldsymbol{\beta} \| + E[Y^2],$$
where, in the second-to-last step, the Cauchy-Schwarz inequality was used. As $\|\boldsymbol{\beta}\|\rightarrow +\infty$, then $E[(Y-X' \boldsymbol{\beta})^{2}]\rightarrow +\infty$, that is, the quadratic function $h(\boldsymbol{\beta})=E[(Y-X' \boldsymbol{\beta})^{2}]$ is coercive. The next step in the proof is to transform the original \emph{true} problem (\ref{P}) into an equivalent one with a feasible compact region $\boldsymbol{B}^*$. Given $\boldsymbol{\beta}^*\in \boldsymbol{B}$, since $h(\boldsymbol{\beta})=E[(Y-X'\boldsymbol{\beta})^{2}]$ is coercive, there exists $R$ such that $$E[( Y - X'\boldsymbol{\beta})^2] \,\,> E[( Y - X'\boldsymbol{\beta}^*)^2 + \lambda \| \mathcal{A}\boldsymbol{\beta}^*\|_1],$$ for all $\boldsymbol{\beta}$ with $\| \boldsymbol{\beta}\| > R$. For that reason, the problem (\ref{P}) can be reduced to the feasible compact region
$\boldsymbol{B}^* = \boldsymbol{B}\cap\{\boldsymbol{\beta}: \,\, \|\boldsymbol{\beta}\|  \leq  R\}$, which implies that the optimal solution is reached.

Finally, the uniqueness of solution is a consequence of the strict convexity of the objective function.

\subsection*{Proof of Theorem \ref{consis_theo}}

For the sake of simplicity, $\boldsymbol{\beta}^{CSCLasso}(\lambda)$ and $\boldsymbol{\hat{\beta}}^{CSCLasso}(\lambda)$ will be denoted henceforth by $\boldsymbol{\beta}$  and  $\boldsymbol{\hat{\beta}}$, respectively. In addition, let us consider the nonempty  compact set $C=\boldsymbol{B}\cap\{\boldsymbol{\beta}: \,\, \|\boldsymbol{\beta}\|  \leq  R\}$, where $R$ is chosen according to the proof of Theorem \ref{ExandUniq}.
\par
Theorem \ref{consis_theo} is a direct consequence of \emph{Theorem 5.3} in \cite{shapiro2009lectures} under some technical conditions, namely:
\begin{itemize}
\item[C1.] The expected value function $E[(Y-X'\mathbf{\boldsymbol{\beta}})^{2}+ \lambda \| \mathcal{A}\mathbf{\boldsymbol{\beta}}\|_1]$ is finite valued and continuous on $C$.
\item[C2.] $\dfrac{1}{n}\sum_{i=1}^{n}((y_{i}-x^{'}_{i} \boldsymbol{\beta})^{2}+ \lambda \| \mathcal{A}\boldsymbol{\beta}\|_1)$ converges to
$E[(Y-X'\mathbf{\boldsymbol{\beta}})^{2}+ \lambda \| \mathcal{A}\mathbf{\boldsymbol{\beta}}\|_1]$ w.p.\ 1, as $n\rightarrow\infty$, uniformly in $\boldsymbol{\beta}\in C$.
\end{itemize}

Let us denote $F(\boldsymbol{\beta},(Y,X))=(Y-X'\mathbf{\boldsymbol{\beta}})^{2}+ \lambda \| \mathcal{A}\mathbf{\boldsymbol{\beta}}\|_1$. Then, the previous conditions C1 and C2 are consequences of \emph{Theorem 7.48} in \cite{shapiro2009lectures} provided that
\begin{itemize}
\item[A1.] for any $\boldsymbol{\beta} \in C$, the function $F(\cdot,(Y,X))$ is continuous at $\boldsymbol{\beta}$ for almost every $(Y, X)$,
\item[A2.] the function $F(\boldsymbol{\beta},(Y,X))$, with $\boldsymbol{\beta}\in C$, is dominated by an integrable function,
\item[A3.] the sample is i.i.d.
\end{itemize}

Given $(Y,X)$, the function $(Y-X'\mathbf{\boldsymbol{\beta}})^{2}+ \lambda \| \mathcal{A}\mathbf{\boldsymbol{\beta}}\|_1$ is continuous at $\boldsymbol{\beta}$ for any $\boldsymbol{\beta} \in C$, and therefore A1 is fulfilled. The sample is i.i.d.\ by hypothesis, and thus A3 holds too. Finally, in order to prove A2, it is necessary to find a measurable function $g(Y, X)>0$ such that $E[g(Y,X)]< \infty$ and, for every $\boldsymbol{\beta}\in C$, $\mid F(\boldsymbol{\beta}, (Y,X))\mid \leq g(Y,X)$ w.p.\ 1.
Using the Cauchy-Schwarz inequality, one has,
$$\mid F(\boldsymbol{\beta}, (Y, X))\mid=\mid (Y-X\boldsymbol{\beta})^2 + \lambda\|\mathcal{A}\boldsymbol{\beta}\|_1\mid=$$
$$= \mid Y^2 - 2YX'\boldsymbol{\beta} + \boldsymbol{\beta}'XX'\boldsymbol{\beta} + \lambda\|\mathcal{A}\boldsymbol{\beta}\|_1\mid \leq$$
$$\leq Y^2 + (X'\boldsymbol{\beta})^2 + 2\mid YX'\boldsymbol{\beta}\mid + \lambda\|\mathcal{A}\boldsymbol{\beta}\|_1 = $$
$$=Y^2  + \mid X'\boldsymbol{\beta}\mid^2 + 2\mid YX'\boldsymbol{\beta}\mid + \lambda\|\mathcal{A}\boldsymbol{\beta}\|_1 \leq$$
$$ \leq Y^2 + \|X\|^2 \|\boldsymbol{\beta}\|^2 + 2 \|YX\|\|\boldsymbol{\beta}\| + \lambda\|\mathcal{A}\boldsymbol{\beta}\|_1.$$
Let $M_1$ and $M_2$ be given by
\[
M_1 = \max_{\boldsymbol{\beta}\in C}  \|\boldsymbol{\beta}\| \qquad\qquad M_2 = \max_{\boldsymbol{\beta}\in C} |\mathcal{A}\boldsymbol{\beta}|
\]
which are well defined due to the compactness of $C$. Therefore, $g$ can be chosen as
$$g(Y,X)= Y^2 + M_1^2\|X\|^2 + 2M_1 \|YX\| + \lambda M_2,$$ which is positive and, since $E(\|X\|^2)<\infty$, $E(Y^2)<\infty$, $E(\|YX\|)<\infty$, its expected value is finite. In consequence, A2 holds and the proof is concluded.

\newpage
\section*{Appendix: further results}
\hspace*{-3cm}\begin{figure}[h!]
\caption{Heat maps of $\boldsymbol{\hat{\beta}}^{CSCLasso}(\lambda) = (\hat{\beta}^{CSCLasso}_1(\lambda),\ldots,\hat{\beta}^{CSCLasso}_8(\lambda))$ using {\tt prostate} dataset }
\begin{tabular}{llll}
\hspace*{-1cm}\includegraphics[width=6.5cm]{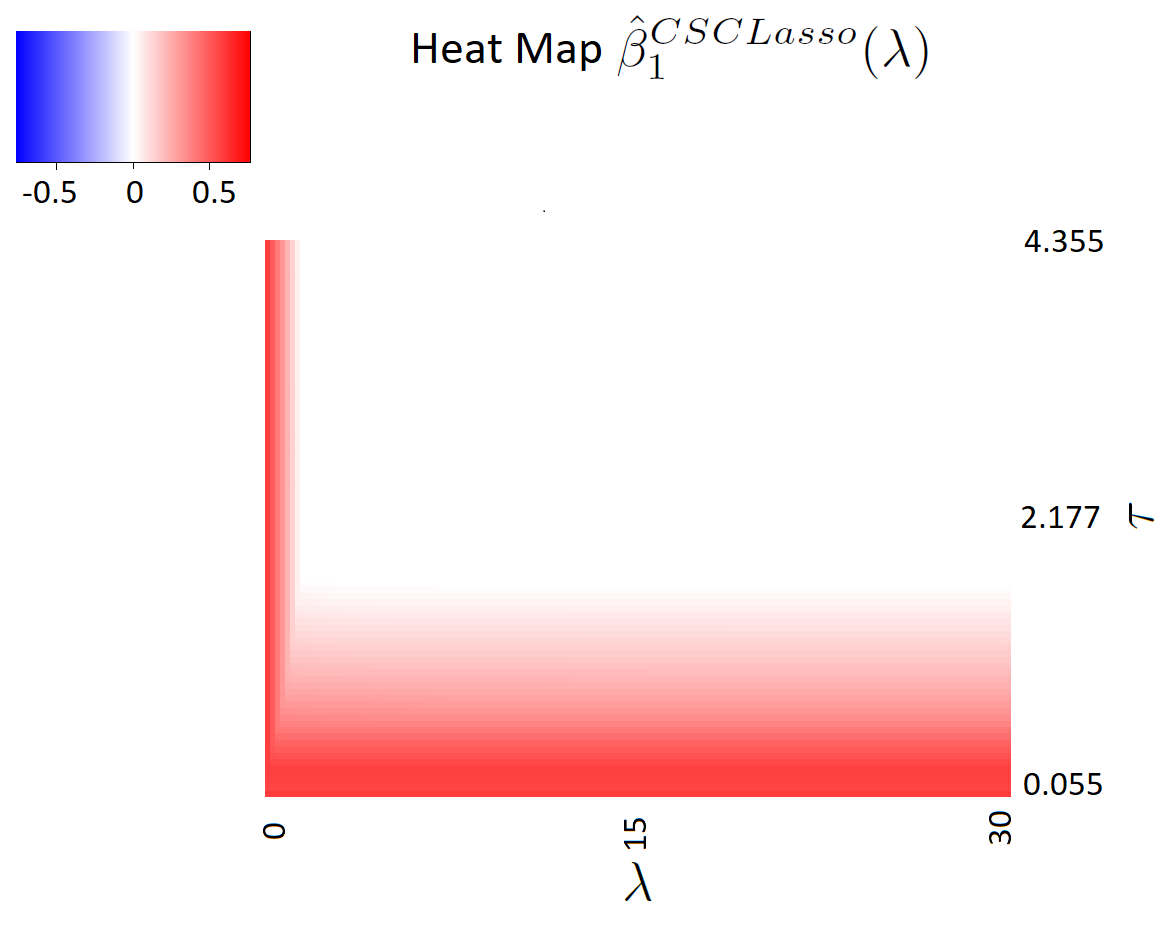}
& \hspace*{-1cm}\includegraphics[width=6.5cm]{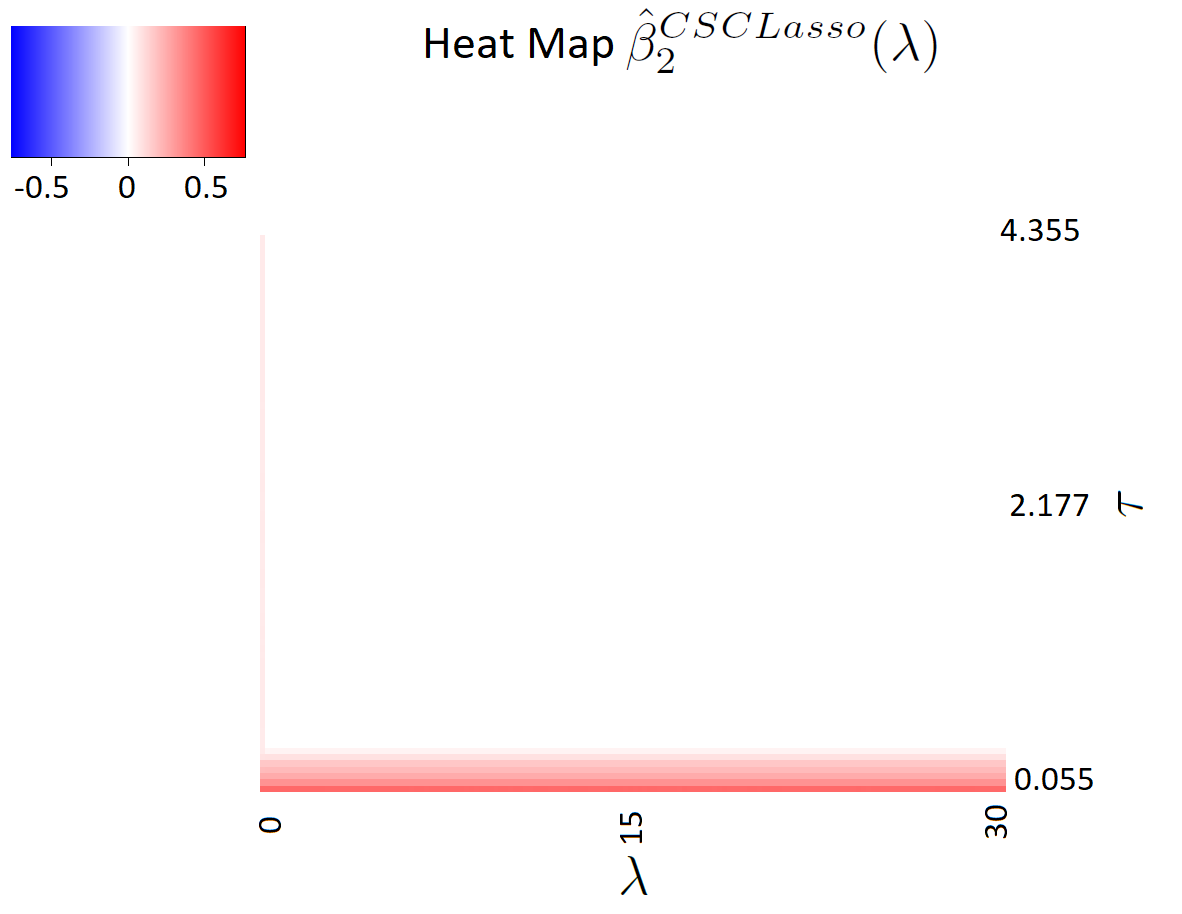}
 & \hspace*{-1cm} \includegraphics[width=6.5cm]{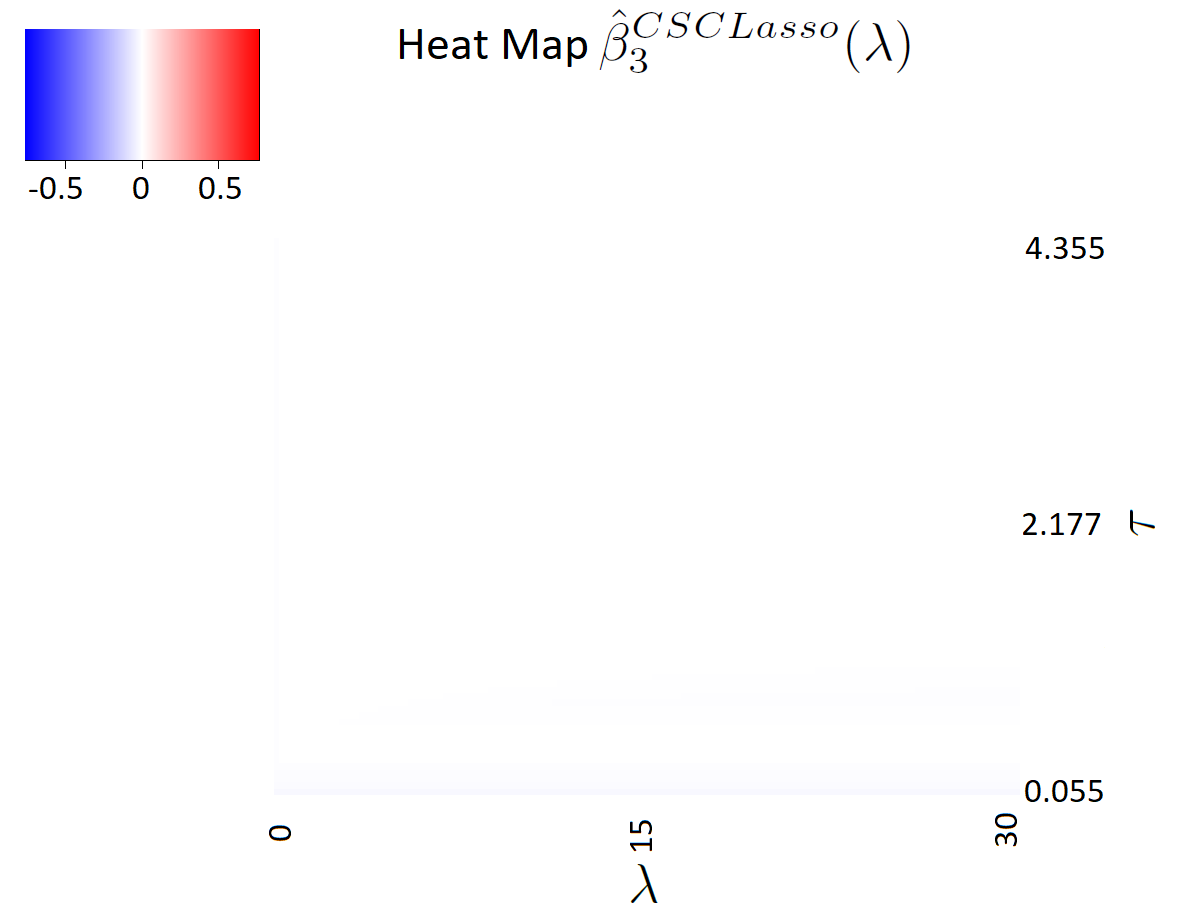}\\
 \hspace*{-1cm}\includegraphics[width=6.5cm]{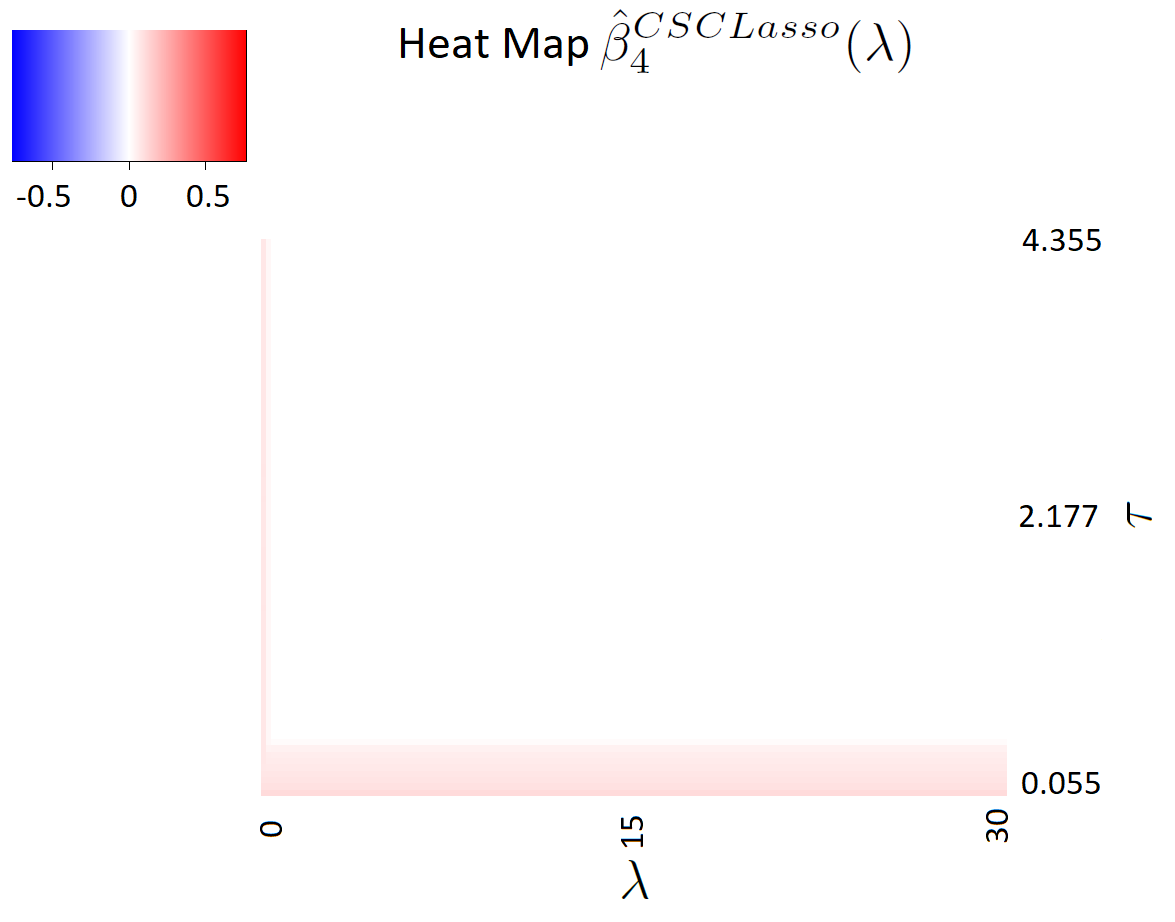}
&\hspace*{-1cm}\includegraphics[width=6.5cm]{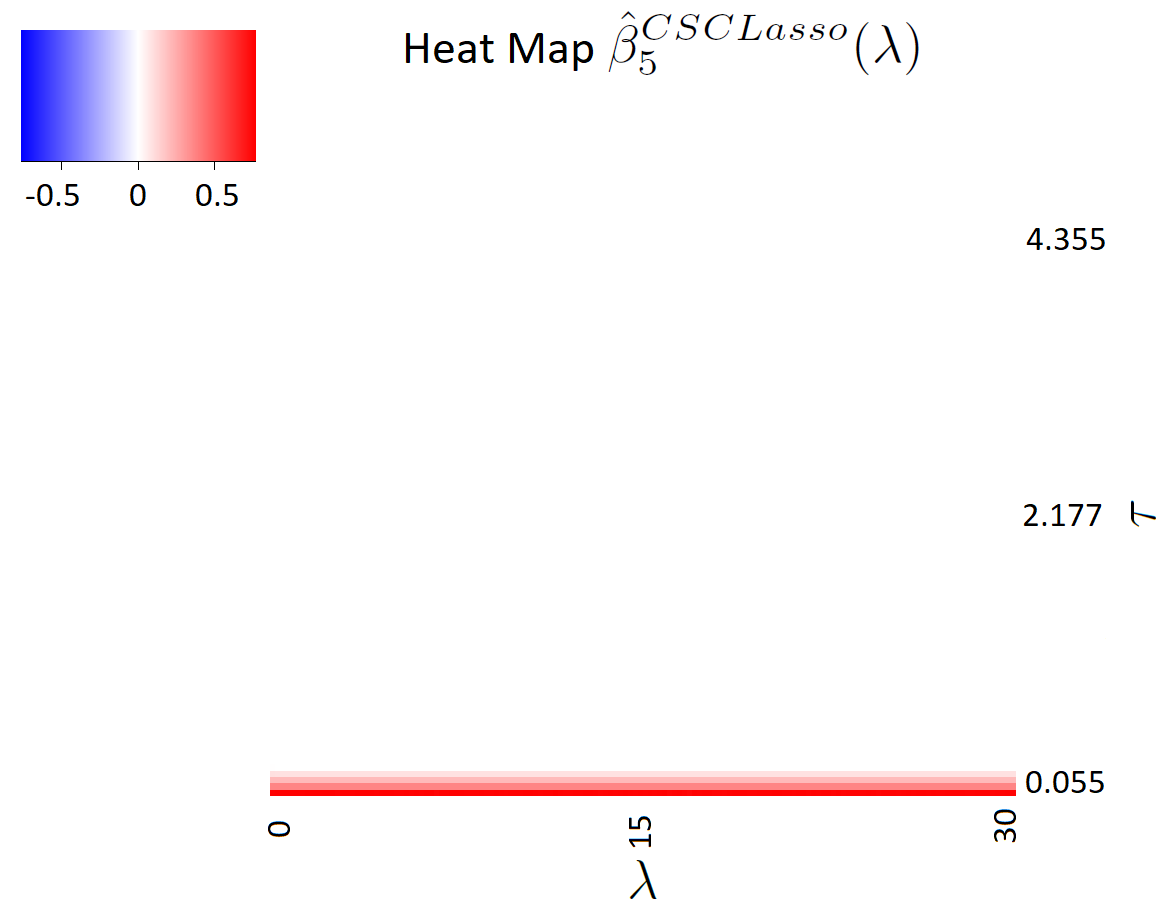}
& \hspace*{-1cm}\includegraphics[width=6.5cm]{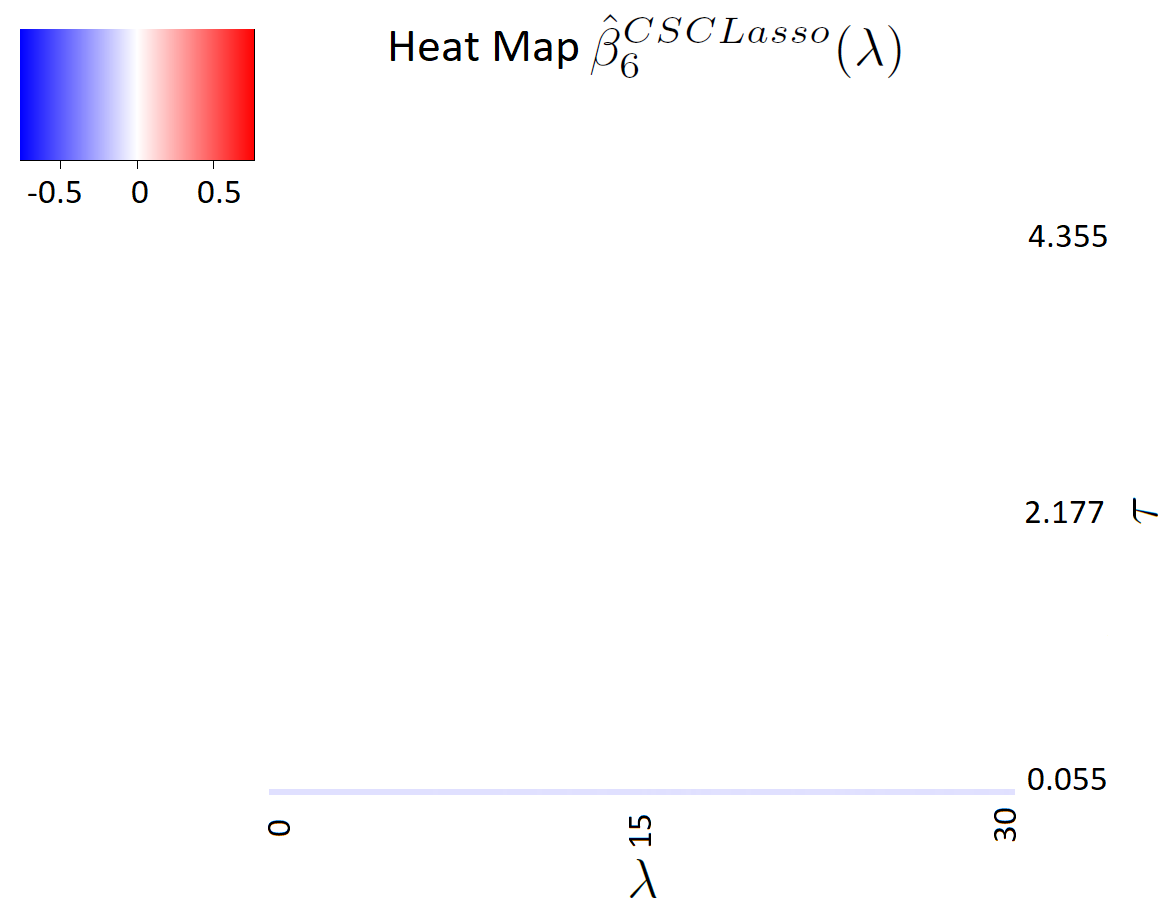}\\
 \hspace*{-1cm}\includegraphics[width=6.5cm]{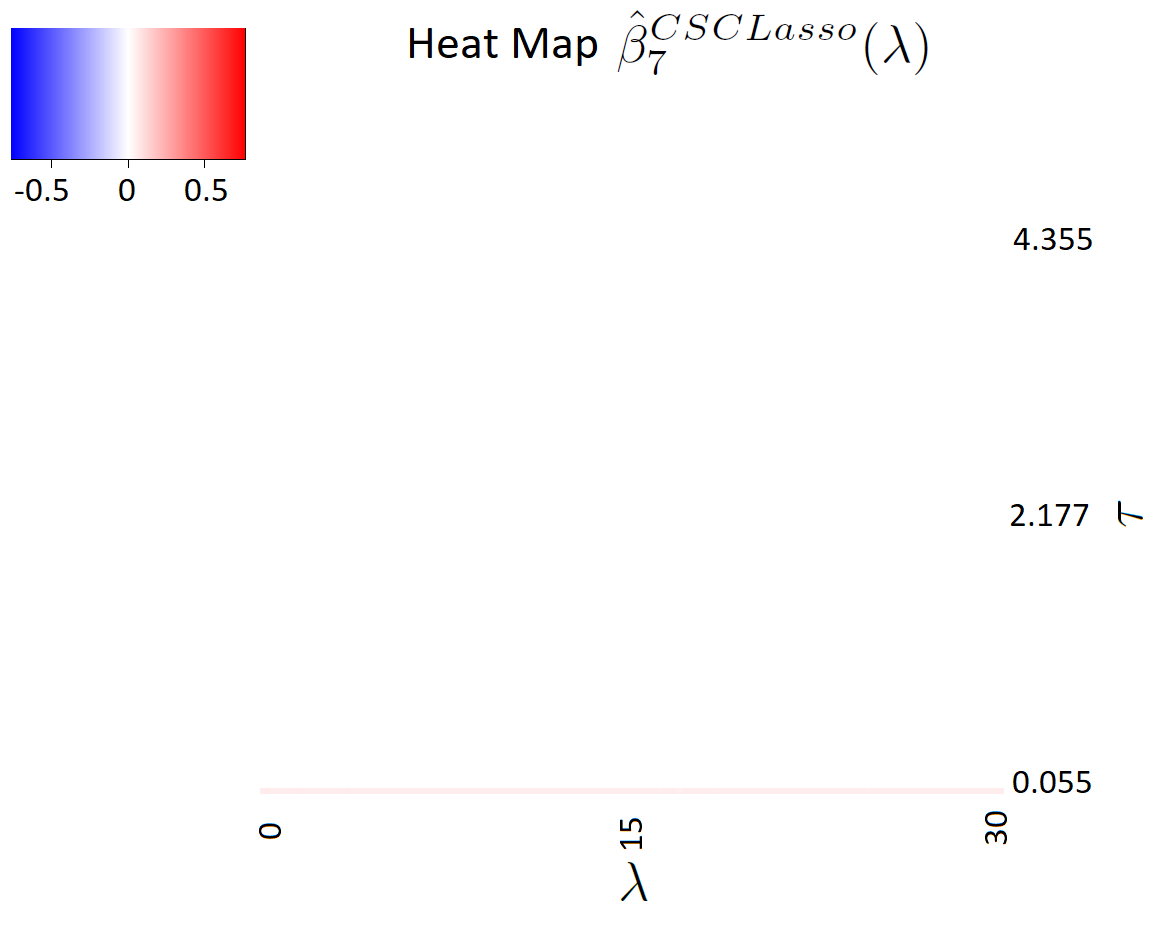}
& \hspace*{-1cm}\includegraphics[width=6.5cm]{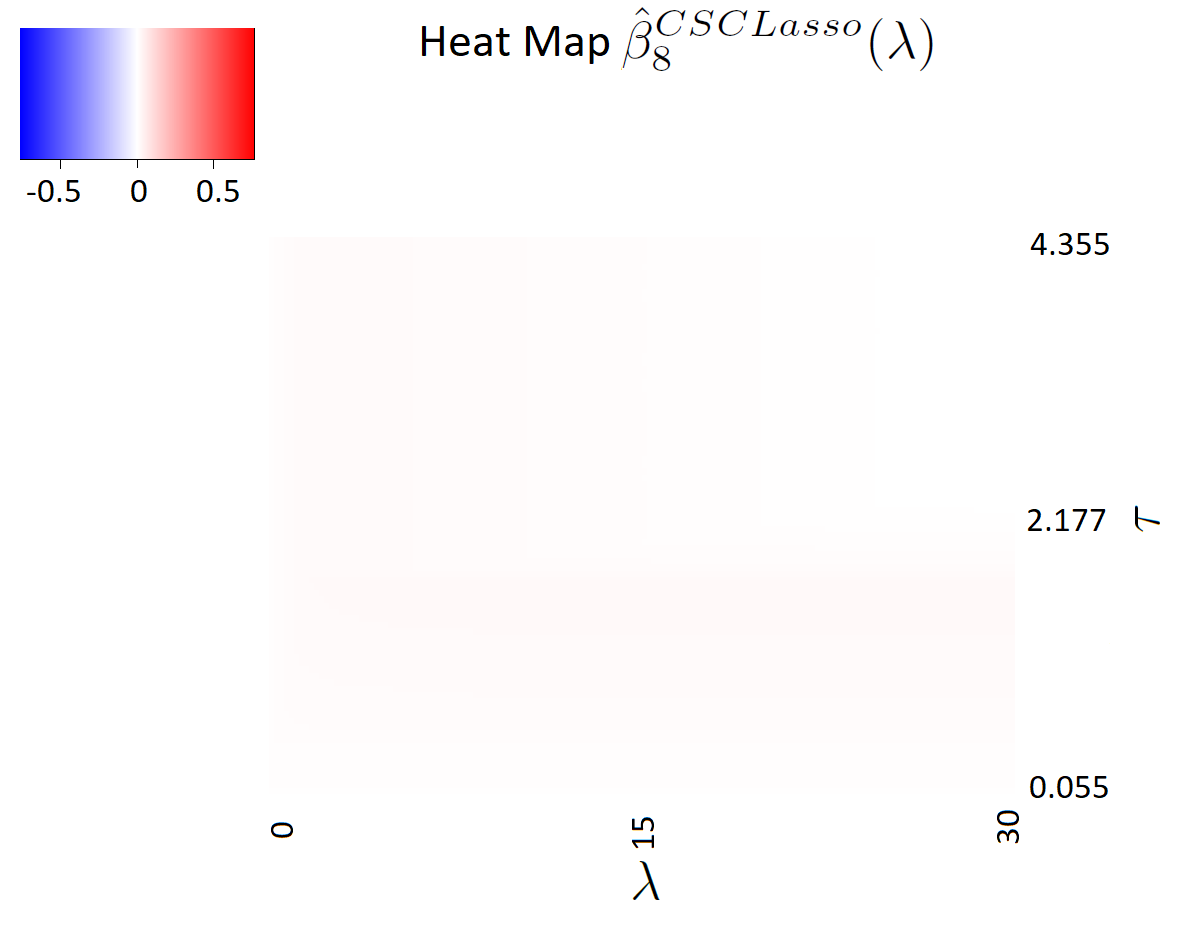}\\
\end{tabular}
\label{fig:Summarizing graphic}
\end{figure}

\begin{figure}[H]
\centering \small
\subfigure{\includegraphics[width=130mm]{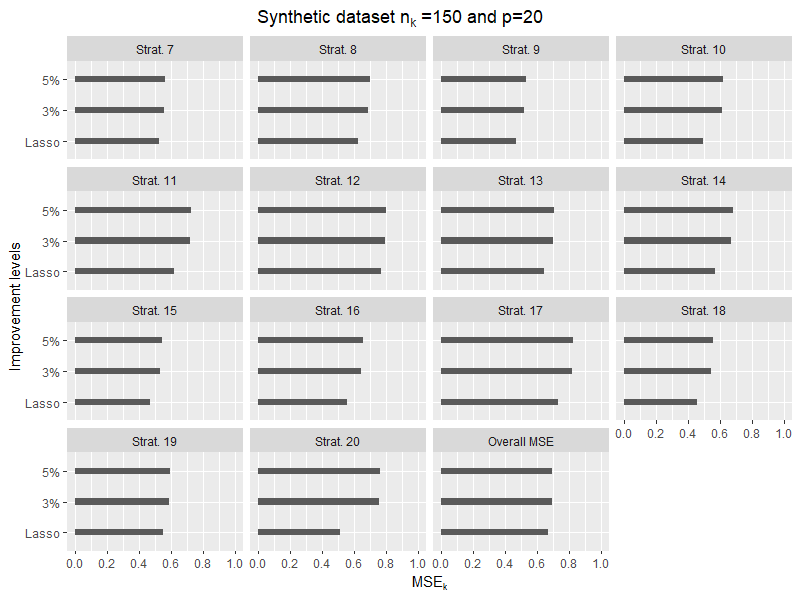}}
\subfigure{\includegraphics[width=130mm]{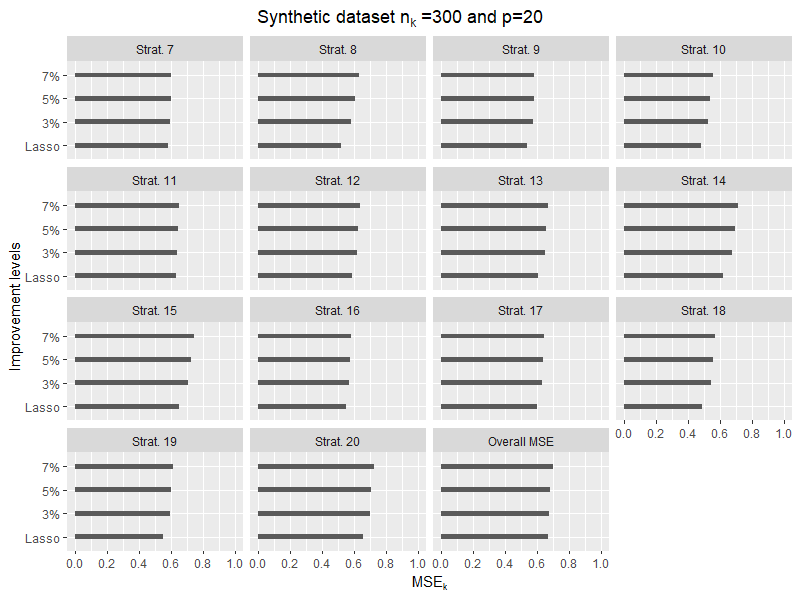}}
\caption{Median $MSE_k$ over the test sets for $k=7,\ldots,20$ under $p=20$ features and the two $n_k$ options. Each subgraph represents one group and the Y-axis shows the different percentages of improvement}
\label{fig:p20}
\end{figure}

\begin{figure}[H]
\centering \small
\subfigure{\includegraphics[width=130mm]{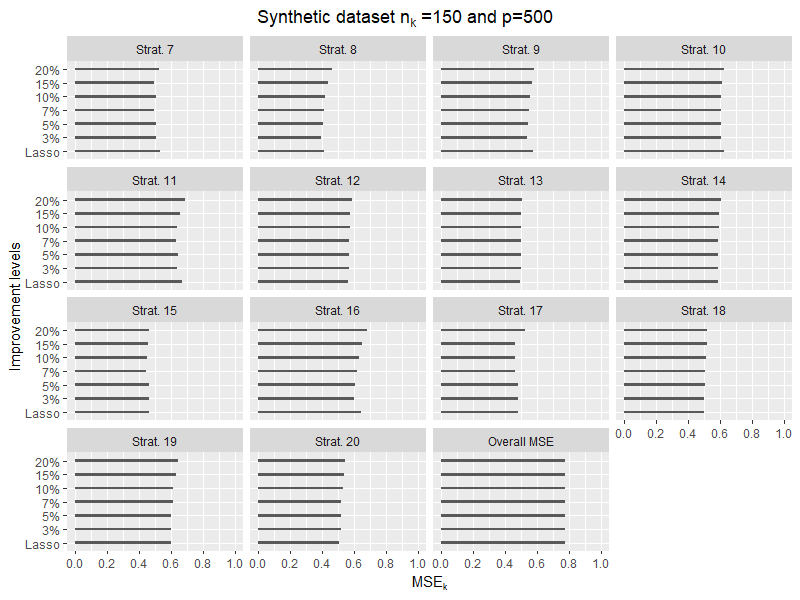}}
\subfigure{\includegraphics[width=130mm]{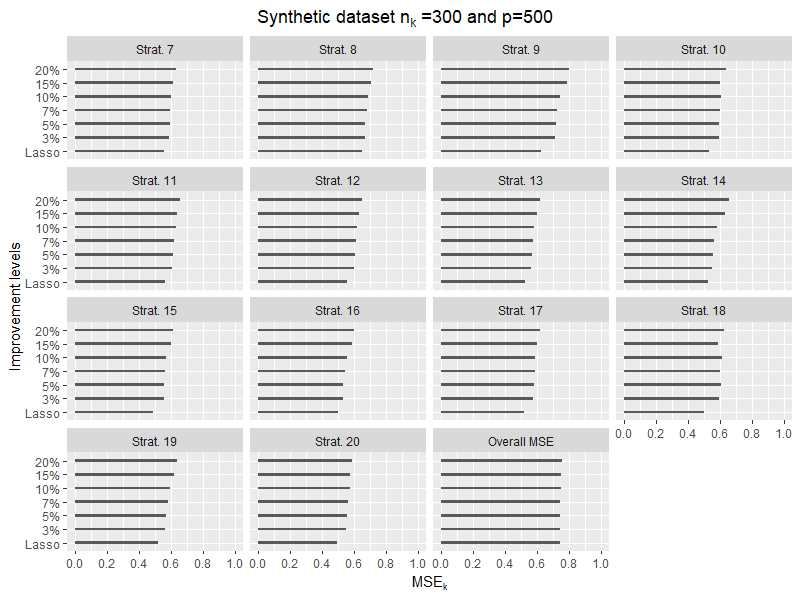}}
\caption{Median $MSE_k$ over the test sets for $k=7,\ldots,20$ under $p=500$ features and the two $n_k$ options. Each subgraph represents one group and the Y-axis shows the different percentages of improvement}
\label{fig:p500}
\end{figure}

\begin{figure}[H]
\centering
\subfigure{\includegraphics[width=150mm]{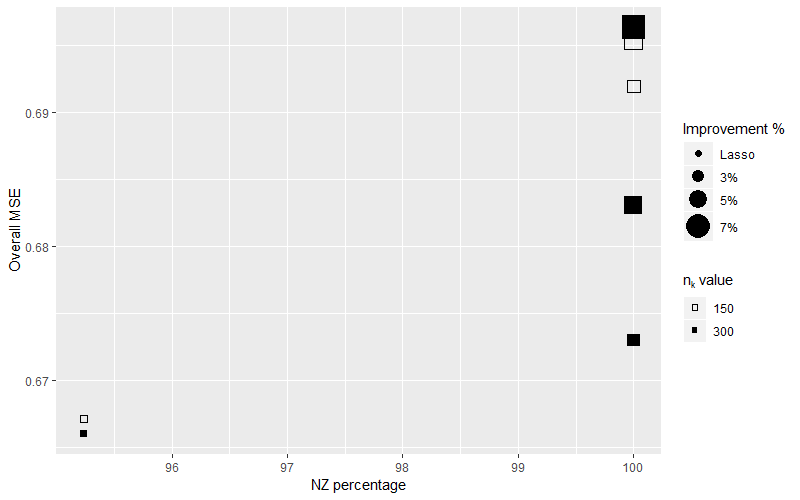}}
\subfigure{\includegraphics[width=150mm]{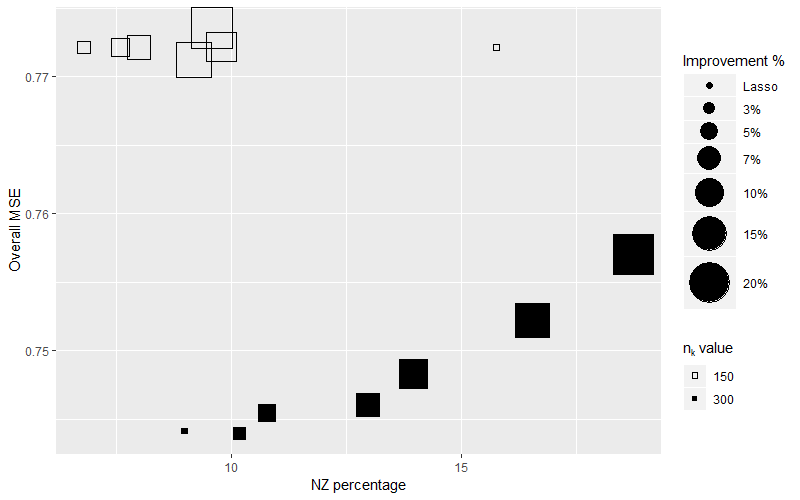}}
\caption{Median overall MSE over the test sets and NZ percentage under the choice $p=20$ (top) and $p=500$ (bottom)}
\label{fig:NZp20and500}
\end{figure}

\begin{figure}[h!]
\centering
\subfigure{\includegraphics[width=59mm]{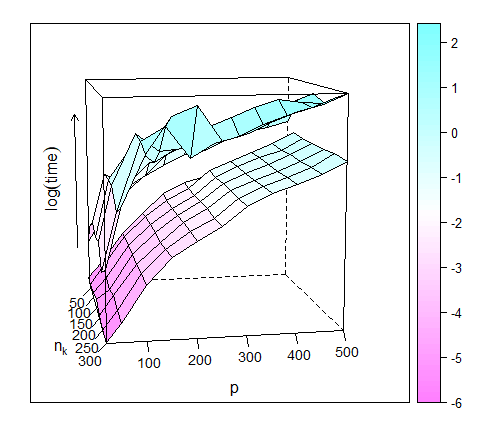}}
\subfigure{\includegraphics[width=59mm]{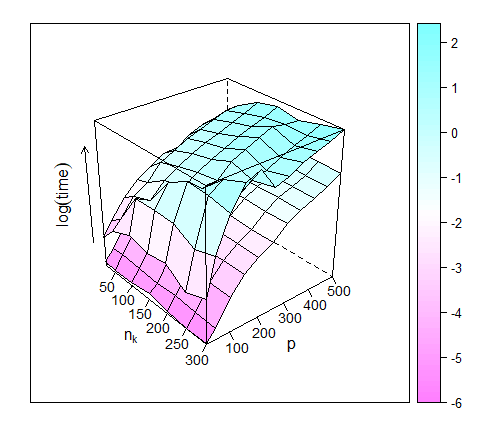}}
\subfigure{\includegraphics[width=59mm]{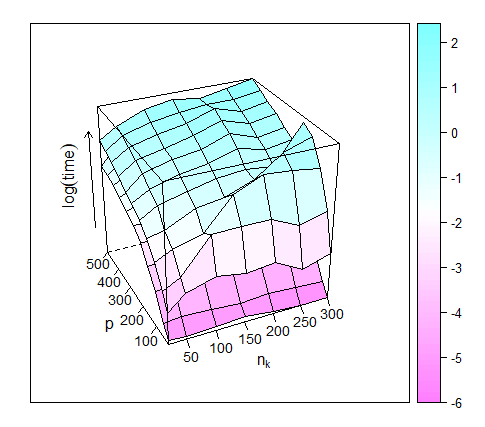}}
\subfigure{\includegraphics[width=59mm]{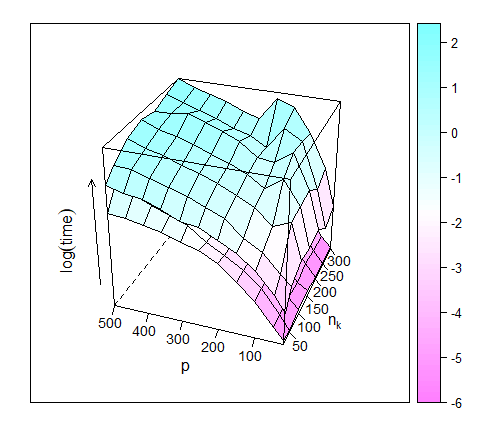}}
\caption{Four perspectives of the logarithm of the user times in seconds for Lasso (bottom surface in the four graphics) and CSCLasso (top surfaces) models across a grid in $n_k$ and $p$}
\label{fig:times}
\end{figure}

To fully understand how the computation time behaves depending on $n_k$ and $p$ values, a grid in both parameters have been inspected. Figure \ref{fig:times} displays the logarithm of the user times in seconds obtained under Lasso and CSCLasso models when $n_k$ and $p$ change. The perspective drawn in the top left figure shows that Lasso model (bottom surface) is solved faster and in a smoother way. Besides, whereas smaller times are obtained for both methods when $n_k$ and $p$ are small, the biggest times are associated to $n_k=300$ and $p=500$.

\bibliographystyle{abbrv}
\bibliography{Bibliography_CLasso}

\end{document}